\documentclass[twocolumn,aps,superscriptaddress,prb]{revtex4-1}
\usepackage{graphicx}
\usepackage{amssymb}
\usepackage{amsmath}
\usepackage{bm}
\usepackage{color}

\usepackage{braket}
\usepackage{siunitx}
\usepackage[top=30truemm,bottom=30truemm,left=25truemm,right=25truemm]{geometry}

\usepackage{gensymb}

\def\Vec{\mathbf}

\def\lsim{\, \lower -0.3ex \hbox{$<$} \kern -0.75em \lower 0.7ex \hbox{$\sim$} \,}
\def\gsim{\, \lower -0.3ex \hbox{$>$} \kern -0.75em \lower 0.7ex \hbox{$\sim$} \,}

\begin{document}

\title{Topological invariants 
in two-dimensional quasicrystals}
\author{Mikito Koshino and Hiroki Oka}
\affiliation{Department of Physics, Osaka University,  Osaka 560-0043, Japan}
\date{\today}

\begin{abstract} 
We study the topological characterization of the energy gaps in general two-dimensional quasiperiodic systems consisting of multiple periodicities, 
represented by twisted two-dimensional materials.
We show that every single gap is uniquely characterized by a set of integers,
which quantize the area of the momentum space in units of multiple Brillouin zones defined in the redundant periodicities.
These integers can be expressed as the second Chern numbers,
by considering an adiabatic charge pumping under a relative slide of different periodicities,
and using a formal relationship to the four-dimensional quantum Hall effect.
The integers are independent of commensurability of the multiple periods,
and invariant under arbitrary continuous deformations such as a relative rotation of twisted periodicities.
\end{abstract}

\maketitle

\section{Introduction}

The topological phase of matter is a fundamental concept to understand the quantum properties in crystalline solids.
The topological classification relies on the existence of a gap in the energy spectrum,
where the topological invariant is determined by the dependence of eigenstates below the gap on the Bloch wave number.
\cite{ryu2010topological,qi2011topological}
The typical example is the integer quantum Hall effect, where the quantized Hall conductivity is expressed by the first Chern number, or an integral of the Berry curvature of the occupied states over the entire Brillouin zone. \cite{thouless1982quantized,kohmoto1985topological}

The topological properties of quasicrystalline systems have also attracted much attention.
Quasiperiodic systems can also have energy gaps \cite{zoorob2000complete,kaliteevski2000two,dyachenko2007three,krajvci2007topologically}, 
while the lack of the Bloch bands makes the definition of topological numbers a nontrivial problem.
Several theoretical works have been devoted to topological characterization 
of various one-dimensional (1D) \cite{lang2012edge,mei2012simulating,kraus2012topological,kraus2012topologicalequivalence,satija2013chern,ganeshan2013topological,verbin2013observation,verbin2015topological,lohse2016thouless,marra2020topologically,zilberberg2021topology,yoshii2021charge}
and two-dimensional (2D) quasiperiodic systems \cite{kraus2013four,tran2015topological,bandres2016topological,cain2020layer,rosa2021topological,fujimoto2020topological,zhang2020topological,su2020topological}.
In particular, one-dimensional quasicrystals are characterized by the adiabatic charge pumping,
where the number of the transfered charge under a relative slide of a single periodic structure to the other
is given by the first Chern number, in an analogous manner to the quantum Hall effect. \cite{thouless1983quantization,niu1986quantum,kraus2012topological,fujimoto2020topological}

Recent developments in the study of 2D materials 
gave rise to a new class of 2D quasicrystals controlled by twist. \cite{lopes2007graphene,mele2010commensuration,trambly2010localization,shallcross2010electronic,
morell2010flat,bistritzer2011moirepnas,moon2012energy,de2012numerical,dean2010boron,ponomarenko2013cloning,hunt2013massive,cao2018unconventional,cao2018mott,zou2018band,koshino2018maximally,balents2020superconductivity}
When two atomic layers are overlaid with an arbitrary rotation angle, the periodicities of the individual layers 
do not generally match, and the entire system becomes quasi-periodic.
A remarkable feature of these twisted 2D quasicrystals is that the electronic structure can be continuously modified
by changing the twist angle or any deformations of the individual lattice structure.
For instance, the twisted bilayer graphene has a highly-tunable band structure depending on the twist angle,
ranging from the moir\'{e} flat bands in the small angle regime\cite{lopes2007graphene,mele2010commensuration,trambly2010localization,shallcross2010electronic,
morell2010flat,bistritzer2011moirepnas,moon2012energy,de2012numerical,cao2018unconventional,cao2018mott,zou2018band,koshino2018maximally},  
to a 12-fold rotationally symmetric quasicrystal at 30$^\circ$.\cite{stampfli,
ahn2018dirac,yao2018quasicrystalline,moon2019quasicrystalline,crosse2020quasicrystalline,ha2021macroscopically}

Then, one can ask, what is a topological number to characterize energy gaps in a quasiperiodic system, which is invariant under a continuous 
structural deformation such as a relative rotation? 
In our previous work \cite{oka2021fractal}, we studied the energy spectrum in a quasiperiodic system of graphene sandwiched by hexagonal boron nitride (hBN)
\cite{finney2019tunable,wang2019new,wang2019composite,yang2020situ,onodera2020cyclotron,kuiri2021enhanced,andjelkovic2020double,leconte2020commensurate},
and we found that each energy gap in the spectrum is uniquely characterized by a set of integers,
which quantize the area of a quasi Brillouin zone in the momentum space.
These integers, which we refer to as zone quantum numbers,  were shown to be invariant under interlayer rotation as long as the gap remains open.
It implies that there should be an underlying  topological mechanism which guarantees the quantization of the momentum space area,
while the actual topological expression of the zone quantum numbers is yet to be cleared.

\begin{figure}
\centering
\includegraphics[width=0.8\linewidth]{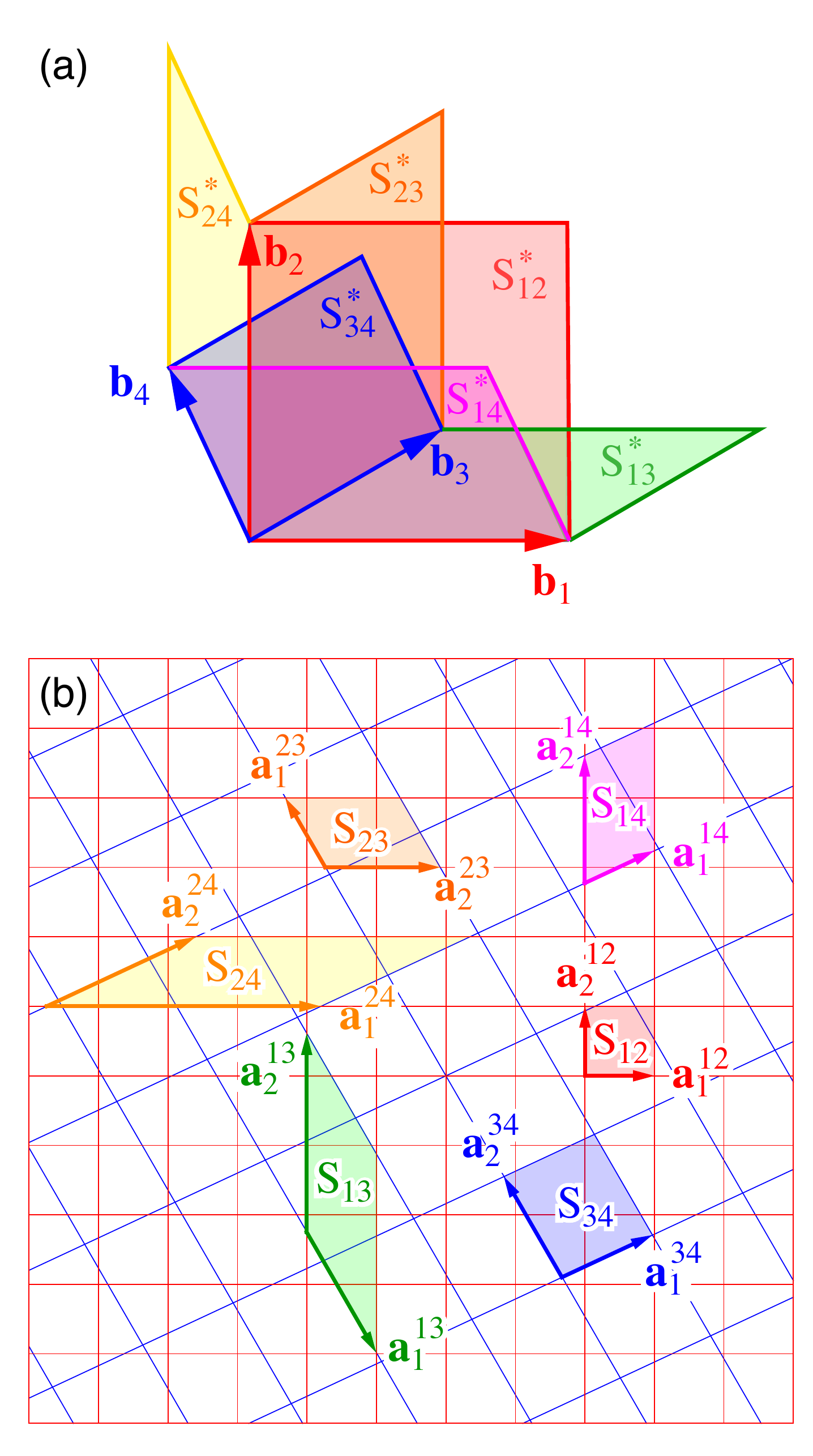}
\caption{(a)  Momentum-space unit areas $S^*_{ij} = (\Vec{b}_i \times \Vec{b}_j)_z$.
(b)  Real-space unit areas $S_{ij}$ and the associated lattice vectors $\Vec{a}^{ij}_1$ and $\Vec{a}^{ij}_2$.
The red (blue) grid lines represent the the wave surfaces of $e^{i \Vec{b}_1\cdot\Vec{r}}$ and $e^{i \Vec{b}_2\cdot\Vec{r}}$
($e^{i \Vec{b}_3\cdot\Vec{r}}$ and $e^{i \Vec{b}_4\cdot\Vec{r}}$).
}
\label{fig_unit_areas}
\end{figure}

In this paper, we consider general 2D doubly-periodic systems (i.e., 2D Hamiltonian with periodic potentials 
$V^\alpha(\Vec{r})$ and $V^\beta (\Vec{r})$ which are generally incommensurate) and 
show that six zone quantum numbers associated with  every single energy gap turn out to be the second Chern numbers.
The zone quantum numbers are simply defined as follows.
The doubly-periodic system has the redundant reciprocal lattice vectors 
$\Vec{b}_1, \cdots, \Vec{b}_4$ ($\Vec{b}_1, \Vec{b}_2$ of $V^\alpha$, and 
$\Vec{b}_3, \Vec{b}_4$ of $V^\beta$) on 2D plane,
where we can define six fundamental Brillouin zones $S^*_{ij} =  (\Vec{b}_i \times \Vec{b}_j)_z$ by taking cross product of two distinct vectors
[Fig.\ \ref{fig_unit_areas}(a)]. Each energy gap is then characterized by six zone quantum numbers $\nu_{ij}$ 
such that the electron density below the gap is quantized as $n_e=\sum_{\langle i,j \rangle} \nu_{ij} S^*_{ij} /(2\pi)^2$.
The associated momentum area $\sum_{\langle i,j \rangle} \nu_{ij} S^*_{ij} $ corresponds to a quasi Brillouin zone, which is a 
single polygon composed of the composite Bragg planes given by $\Vec{b}_1, \cdots, \Vec{b}_4$, and is
a line of the gap opening momenta on the free electron band in the infinitesimal potential limit. \cite{oka2021fractal}

We verify the equivalence between the zone quantum numbers $\nu_{ij}$  and the second Chern numbers
by considering an adiabatic charge pumping under a relative slide of a periodic potential to the other
(corresponding to interlayer sliding in twisted 2D materials).
First, we demonstrate that the number of pumped charge is directly expressed in terms of the zone quantum numbers,
by a simple argument using the charge continuity in a potential deformation.
Then we describe the same pumping process in an alternative approach using the dimensional reduction of the four-dimensional (4D)  quantum Hall effect,
\cite{kraus2013four,qi2008topological,price2015four,price2016measurement,lohse2018exploring,zilberberg2018photonic,lu2018topological} 
where the charge pumping is expressed by the second Chern numbers.
By comparing the corresponding equations in the two different approaches, 
we find that the zone quantum numbers are equivalent to the second Chern numbers.
These integers are fixed in a continuous deformation of the potential,
and do not depend on the commensurability of the multiple periodicities.

The systematic characterization of energy gaps presented in this work
would be applicable to general quasicrystaline systems having redundant reciprocal vectors more than the spatial dimension.  
The identification of the topological numbers in quasicrystalline systems brides the fields of quasicrystal and the topological condensed matter physics.

The paper is organized as follows.
In Sec.\ \ref{sec_char}, we present the formulation of the zone quantum numbers.
We calculate the energy spectrum of a twisted double triangular potential as an example, and
identify the zone quantum numbers and the quasi Brillouin zones.
In Sec.\ \ref{sec_pump},  we consider the adiabatic charge pumping under a relative sliding of the double potential,
and show that the pumping charge is quantized by the zone quantum numbers.
In Sec.\ \ref{sec_4D}, we present an alternative approach to describe the adiabatic charge pumping 
using the 4D quantum Hall effect, and find the equivalence between the zone quantum numbers and the second Chern numbers.
A brief conclusion is given in Sec.\ \ref{sec_concl}.

\section{Zone quantum numbers}
\label{sec_char}

\subsection{General formulation}
\label{sec_general}

We consider a doubly-periodic 2D Hamiltonian 
\begin{equation}
H = \frac{\Vec{p}^2}{2m} + V^\alpha(\Vec{r}) + V^\beta (\Vec{r}),
\label{eq_H_2D}
\end{equation}
where $V^\lambda(\Vec{r})\, (\lambda=\alpha, \beta)$ is a periodic potential given by
\begin{equation}
V^\lambda(\Vec{r}) = \sum_{m_1,m_2} V^\lambda_{m_1,m_2} \, e^{i(m_1 \Vec{b}^\lambda_1 + m_2 \Vec{b}^\lambda_2)\cdot\Vec{r}},
\label{eq_H_2D_V}
\end{equation}
and $\Vec{b}_1^{\lambda}, \Vec{b}_2^{\lambda}$ are its primitive reciprocal lattice vectors.
The real-space lattice vectors $\Vec{a}_1^{\lambda}, \Vec{a}_2^{\lambda}$ 
are defined such that $\Vec{a}^{\lambda}_{\mu}\cdot \Vec{b}^{\lambda}_{\nu} = 2\pi \delta_{\mu\nu}$.
It is useful to introduce serial indexes to label the four reciprocal lattice vectors as
\begin{align}
&(\Vec{b}_1,\Vec{b}_2,\Vec{b}_3,\Vec{b}_4) = (\Vec{b}_1^{\alpha}, \Vec{b}_2^{\alpha},\Vec{b}_1^{\beta}, \Vec{b}_2^{\beta}).
\label{eq_G1234}
\end{align}
We claim that, when the spectrum has an energy gap, the electron density below an energy gap is quantized 
\begin{equation}
n_e= \frac{1}{(2\pi)^2}\sum_{\langle i,j \rangle} \nu_{ij} S^*_{ij}  =  \sum_{\langle i,j \rangle} \frac{\nu_{ij}}{S_{ij}}.
\label{eq_ne_2D}
\end{equation}
Here $\nu_{ij} (i,j = 1,2,3,4)$ are zone quantum numbers which characterize the gap, and 
$\langle i,j \rangle$ represents a pair of different indeces. $S^*_{ij}$ and $S_{ij}$ are defined by
\begin{align}
&S^*_{ij} = (\Vec{b}_i \times \Vec{b}_j)_z, \quad S_{ij} = (2\pi)^2/S^*_{ij},
\label{eq_S_star}
\end{align} 
where $(\cdots)_z$ represents the $z$-component perpendicular to the plane.
$S^*_{ij}$ is a momentum space area spanned by two distinct reciprocal lattice vectors chosen from $\Vec{b}_1,\Vec{b}_2,\Vec{b}_3,\Vec{b}_4$,
and $S_{ij}$ is its real-space counterpart.
We have six independent areas $S^*_{12}, S^*_{13}, S^*_{14}, S^*_{23}, S^*_{24}, S^*_{34}$ as illustrated in Fig.\ \ref{fig_unit_areas}(a),
and we have $S^*_{ji}=-S^*_{ij}$ and $S^*_{ii}=0$ from the definition. 
Accordingly, we have six zone quantum numbers $\nu_{12}, \nu_{13}, \nu_{14}, \nu_{23}, \nu_{24}, \nu_{34}$,
and we define $\nu_{ji} = -\nu_{ij}$ and $\nu_{ii}=0$ for consistency.
The areas $S^*_{ij}$ can be regarded as the projection of faces of four-dimensional hypercube onto the physical 2D plane.

The $S_{ij}$ is the area of the parallelogram formed by the wave surfaces of $e^{i \Vec{b}_i\cdot\Vec{r}}$ and $e^{i \Vec{b}_j\cdot\Vec{r}}$,
as shown in Fig.\ \ref{fig_unit_areas}(b). 
For later convenience, we define the lattice vectors 
\begin{equation}
\Vec{a}^{ij}_1 = \frac{S_{ij}}{2\pi} \, (\Vec{b}_j \times \Vec{e}_z) ,\,  \Vec{a}^{ij}_2 = -\frac{S_{ij}}{2\pi}\, (\Vec{b}_i \times \Vec{e}_z),
\label{eq_avecs}
\end{equation}
where $\Vec{e}_z$ is the unit vector perpendicular to the 2D plane.
$(\Vec{a}^{ij}_1,\Vec{a}^{ij}_2)$ is the primitive lattice vector set corresponding to
$(\Vec{b}^{ij}_1,\Vec{b}^{ij}_2) \equiv (\Vec{b}_i, \Vec{b}_j)$ in the momentum space, and it spans the unit cell $S_{ij} = (\Vec{a}^{ij}_1\times \Vec{a}^{ij}_2)_z$ as illustrated in Fig.\ \ref{fig_unit_areas}(b).
The lattice vectors of the potential $\alpha$ and $\beta$ are given by 
$\Vec{a}^{\alpha}_\mu=\Vec{a}^{12}_\mu$ and $\Vec{a}^{\beta}_\mu=\Vec{a}^{34}_\mu$.

\subsection{Example: Twisted triangular potentials}
\label{sec_twisted_trig}

In our previous work\cite{oka2021fractal}, we verified the relation Eq.\ (\ref{eq_ne_2D}) in  a double-moir\'e system of graphene sandwiched by hexagonal boron nitride.
There only four areas out of the six $S^*_{ij}$'s are independent due to the intrinsic 120$^\circ$ rotational symmetry,
and we identified the corresponding four zone quantum numbers for the energy gaps in the spectrum.
The complete set of six integers can be obtained by allowing a slight deformation to break the symmetry.
In the following, we demonstrate a full identification of the six characteristic integers in a double period system Eq.\ (\ref{eq_H_2D})
with twisted double triangular potential.
A similar calculation with twisted square potential is presented in Appendix \ref{sec_twisted_sq}.

\begin{figure*}
\centering 
\includegraphics[width=1.\linewidth]{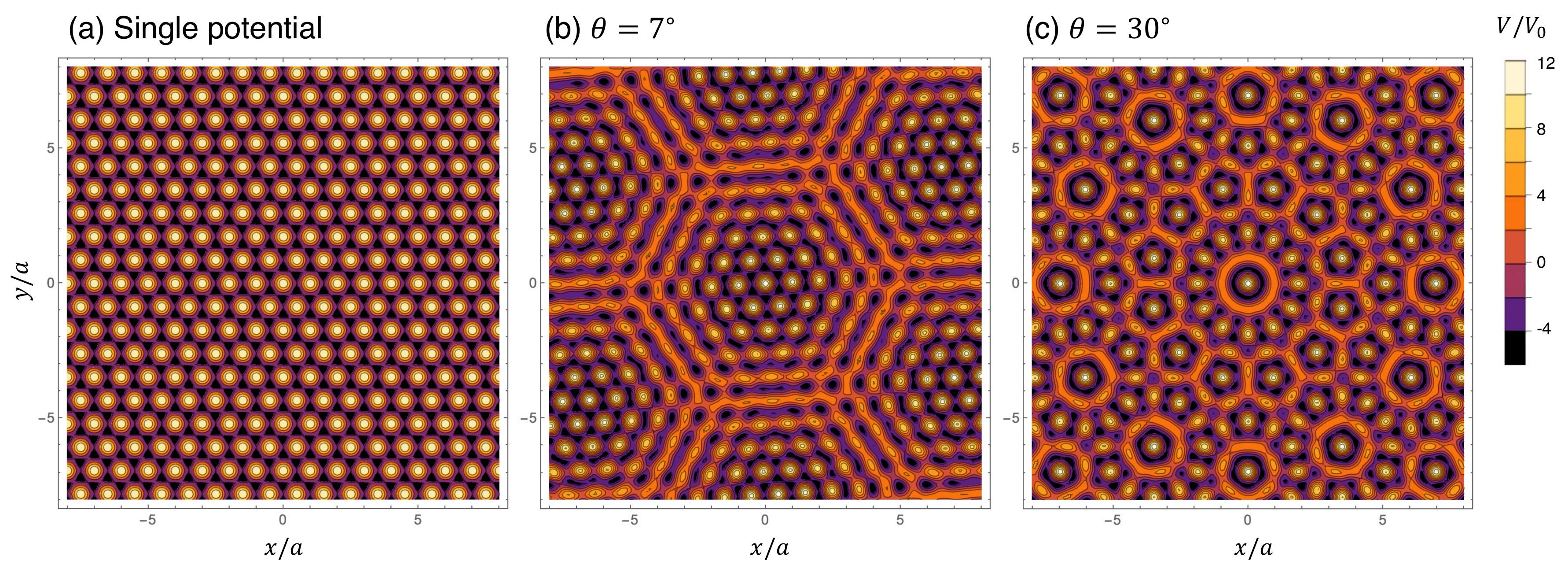}
\caption{Contour plots of  (a) a single triangular potential, and (b) twisted double triangular potentials with $\theta = 7^\circ$ and (c) $\theta = 30^\circ$.
[Eq.\ (\ref{eq_trig_potential})]:
}
\label{fig_twisted_triangular_potential}
\end{figure*}

The twisted double triangular potential is given by
\begin{equation}
V^\lambda(\Vec{r}) = 2V_0 \sum_{\mu=1}^3 \cos[\Vec{b}^\lambda_\mu \cdot (\Vec{r}-\Vec{r}_0^\lambda)],
\label{eq_trig_potential}
\end{equation}
where $\Vec{r}_0^\lambda$ is the origin of the potential $\lambda$.
The reciprocal vectors of $\lambda= \alpha$ are given by
\begin{align}
&\Vec{b}^\alpha_1 = \frac{2\pi}{a}
\begin{pmatrix}
1 \\
-1/\sqrt{3}
\end{pmatrix},
\quad
\Vec{b}^\alpha_2 = \frac{2\pi}{a}
\begin{pmatrix}
0 \\
2/\sqrt{3}
\end{pmatrix},
\nonumber\\
&\Vec{b}^\alpha_3 = -\Vec{b}^\alpha_1-\Vec{b}^\alpha_2,
\label{eq_trig_G}
\end{align}
and those of $\beta$ are defined by 
\begin{align}
\Vec{b}^\beta_\mu= R(\theta) \,\, \Vec{b}^\alpha_\mu,
\label{eq_trig_G_beta}
\end{align}
where $R(\theta)$ is a 2D rotation matrix of angle $\theta$. 
The corresponding primitive lattice vectors are
\begin{align}
&\Vec{a}^\alpha_1 = a
\begin{pmatrix}
1 \\
0
\end{pmatrix},
\quad
\Vec{a}^\alpha_2 = a
\begin{pmatrix}
1/2\\
\sqrt{3}/2
\end{pmatrix},
\nonumber\\
&\Vec{a}^\beta_\mu= R(\theta) \,\, \Vec{a}^\alpha_\mu.
\label{eq_trig_L}
\end{align}
The potential profile is presented in Fig.\ \ref{fig_twisted_triangular_potential},
for (a) a single potential,  (b) double potential with $\theta = 7^\circ$ and (c) $\theta = 30^\circ$.
In the following, the potential amplitude (identical in $\alpha$ and $\beta$) is taken as $V_0 = 0.213\varepsilon_0$,
where $\varepsilon_0 = \hbar^2/(2ma^2)$.

\begin{figure*}
\centering
\includegraphics[width=1.\linewidth]{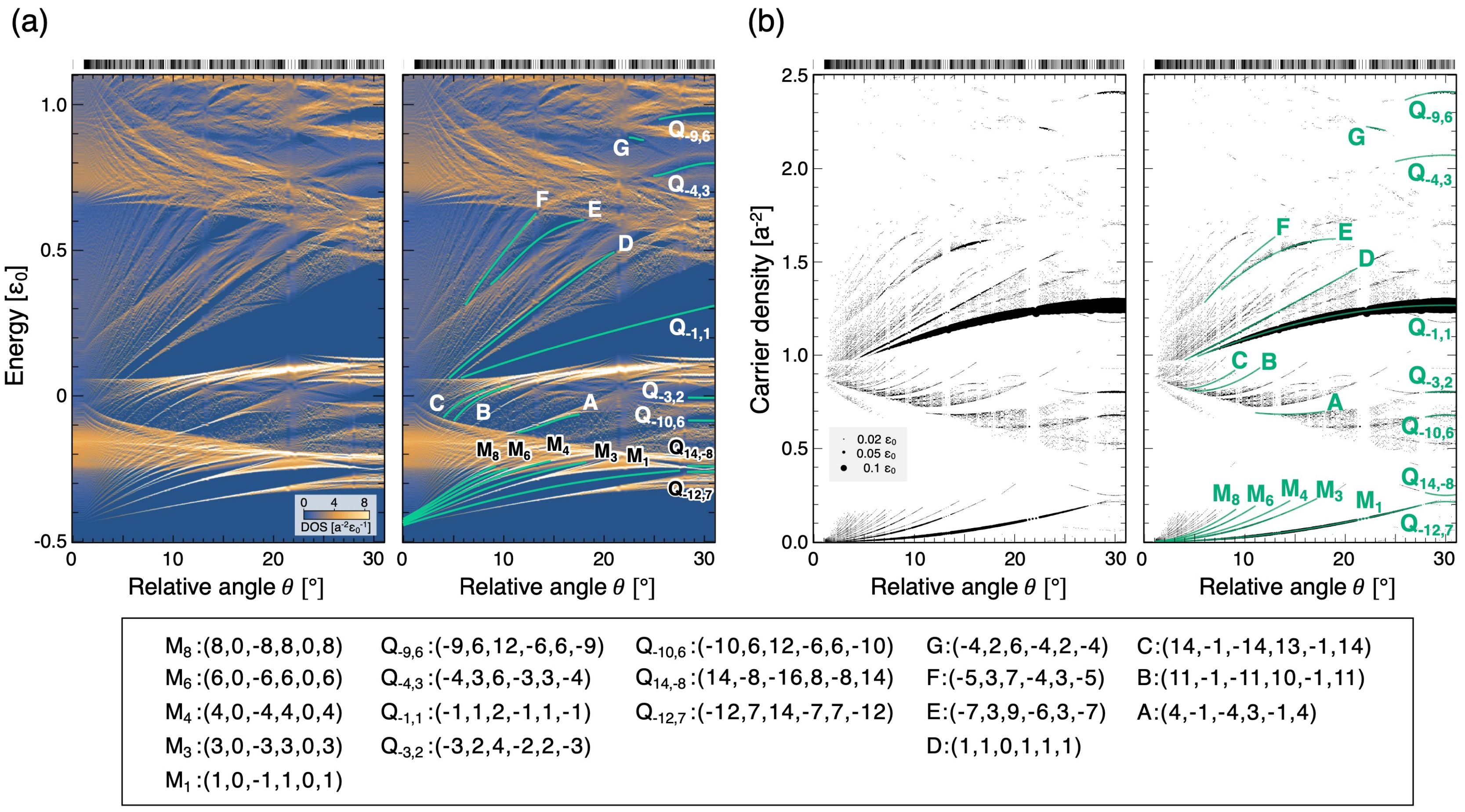}
\caption{(a) Density map of the DOS plotted against $\theta$ and energy in twisted double triangular potentials.
The major gaps are highlighted and labeled in the right panel.
The array of bars in the upper part of the figures represents the commensurate approximants considered in the calculation.
(b) Corresponding plot with vertical axis converted to the electron density, where the dot size indicates the gap width.
The bottom table presents the zone quantum numbers $(\nu_{12}, \nu_{13}, \nu_{14}, \nu_{23}, \nu_{24}, \nu_{34})$ of the highlighted gaps.
}
\label{fig_spectrum_triangular}
\end{figure*}

Generally, the potentials of $\alpha$ and $\beta$ do not have common period and hence the system does not have the global translational symmetry.
Here we calculate the energy spectrum by using commensurate approximants \cite{oka2021fractal},
which are obtained by slightly deforming the periodicity of $\alpha$ or $\beta$,
so that the system has a finite super unit cell with an area $S_{\rm c}$. 
The detail of the commensurate approximant method  is described in Appendix \ref{sec_comm_approx}.
We prepare a series of commensurate approximants to mimic the continuous rotation of the twist angle between the two potentials,
and calculate the energy bands and the density of states (DOS) for all the systems.
For each approximant, we average the DOS over the relative translation $\Delta \Vec{r}_0 = \Vec{r}_0^\alpha - \Vec{r}_0^\beta$
to obtain a continuous spectrum as a function of the twist angle.

Figure \ref{fig_spectrum_triangular}(a) shows the density map of the DOS plotted against $\theta$ and energy,
where the brighter color indicates larger DOS, and the dark blue region represents the gap.
The array of bars in the upper part of the figure represents the commensurate approximants considered in the calculation.
Figure \ref{fig_spectrum_triangular}(b) is the corresponding plot 
with vertical axis converted to the electron density, where dots represent energy gaps with the gap width indicated by the size.

For each energy gap, the zone quantum numbers $\nu_{ij}$ can be identified  in the following manner.
In a commensurate approximant, the momentum space area $S^*_{ij}$'s have the greatest common divisor
$S^*_{\rm c} = (2\pi)^2/S_{\rm c}$, 
and hence they are expressed as $S^*_{ij}  = s_{ij} S^*_{\rm c}$ with integers $s_{ij}$.
Also, the electron density below a given band gap is quantized as $n_e = [S^*_{\rm c}/(2\pi)^2] r$, where
$r$ in the number of the occupied Bloch subbands.
Then, Eq.\ (\ref{eq_ne_2D}) becomes the Diophantine equation $r = \sum_{\langle i,j \rangle} \nu_{ij} s_{ij}$.
By considering more than six commensurate approximants sharing the same energy gap,
we have the Diophantine equations as many as the number of the approximants,
and we finally obtain the integers $\nu_{ij}$ as a unique solution of the set of equations.
It should be also noted that the original double triangular potential, Eq.\ (\ref{eq_trig_potential}), has constraints on  $S^*_{ij}$'s
such as $S^*_{12}=S^*_{34}$ and $S^*_{13}=S^*_{24}$ due to the high spatial symmetry, which prevents the full identification of $\nu_{ij}$'s. 
This problem is removed by including systems with the symmetry slightly broken in the set of commensurate approximants.
The actual calculation of the zone quantum numbers using the commensurate approximants is presented in Appendix \ref{sec_comm_approx}.

In the bottom of Fig.\ \ref{fig_spectrum_triangular}, 
we present the zone quantum numbers $(\nu_{12}, \nu_{13}, \nu_{14}, \nu_{23}, \nu_{24}, \nu_{34})$ 
identified for some major gaps labelled in Fig.\ \ref{fig_spectrum_triangular}(a) and (b).
The series $M_n$ in the low twist angle regime is the moir\'{e} gaps, which has a form of
\begin{equation}
M_n = n (1,0,-1,1,0,1).
\label{eq_M_n}
\end{equation}
In this region, the system is governed by a long-range moir\'e pattern as seen in Fig.\ \ref{fig_twisted_triangular_potential}(b), 
and the discrete levels separated by $M_n$ can be viewed as the Bloch subbands of the moir\'e superlattice.
The reciprocal lattice vectors for the  moir\'e period are given by
\begin{align}
\Vec{G}^{\rm M}_1 = \Vec{b}_1 - \Vec{b}_3, \quad \Vec{G}^{\rm M}_2 = \Vec{b}_2 - \Vec{b}_4,
\label{eq_moire_G}
\end{align}
and the area of the moir\'{e} Brillouin zone becomes
\begin{align}
S^*_{\rm M} = (\Vec{G}^{\rm M}_1 \times \Vec{G}^{\rm M}_2)_z = S^*_{12} - S^*_{14} + S^*_{23} + S^*_{34},
\end{align}
which corresponds to $(1,0,-1,1,0,1)$.
Eq.\ (\ref{eq_M_n}) indicates that the momentum space area is quantized by $S^*_{\rm M}$.

In the large angle region $\theta \gg 1^\circ$,  the long-wavelength picture is no longer valid
and the system cannot be effectively captured by any single periodicity.
At $\theta=30^\circ$, in particular, the system becomes a quasicrystal with 12-fold rotational symmetry \cite{stampfli,
ahn2018dirac,yao2018quasicrystalline,moon2019quasicrystalline,crosse2020quasicrystalline,ha2021macroscopically},
 as shown in Fig.\ \ref{fig_twisted_triangular_potential}(c),
Here we find that the zone quantum numbers always have the form,
\begin{equation}
Q_{m,n} =  (m, n, 2n, -n, n, m).
\label{eq_Q_ab}
\end{equation}
The corresponding electronic density Eq.\ (\ref{eq_ne_2D}) is $n_e = (\sqrt{3}m+3n)/a^2$, 
indicating that there are two distinct units, $\sqrt{3}/a^2$ and $3/a^2$, to quantize the electronic spectrum.

The constraint on the zone quantum numbers Eq.\ (\ref{eq_Q_ab}) is explained as follows.
If we define $\Vec{b}'_i\, (i=1,2,3,4)$ by the 30$^\circ$-rotation of $\Vec{b}_i$,
we have a relation $(\Vec{b}'_1,\Vec{b}'_2,\Vec{b}'_3,\Vec{b}'_4)=(\Vec{b}_3,\Vec{b}_4,\Vec{b}_1+\Vec{b}_2,-\Vec{b}_1)$.
The associated areas $S^*_{ij}{'} = (\Vec{b}'_i \times \Vec{b}_j')_z$
can be expressed by the old areas as $S^*_{12}{'}  = S^*_{34}$, $S^*_{13}{'}  = -S^*_{13}-S^*_{23}$, etc.
When the system is invariant under the 30$^\circ$-rotation,
we should have $\sum_{\langle i,j \rangle} \nu_{ij} S^*_{ij} = \sum_{\langle i,j \rangle} \nu_{ij} S^*_{ij}{'}$
with the identical $\nu_{ij}$.
By using the relationship between $S^*_{ij}{'}$ and $S^*_{ij}$, we obtain constraints for $\nu_{ij}$, and finially find Eq.\ (\ref{eq_Q_ab}).

Other gaps are just labelled as $A,B,C\cdots $ in Fig.\ \ref{fig_spectrum_triangular}.
We see that the zone quantum numbers of any gaps definitely come in a form of $(m, n, r, n-r, n, m)$.
This is explained by the coexistence of the 120$^\circ$ rotational symmetry which requires the form of $(m, n, r, n-r, n, m')$,
and the reflection symmetry with respect to the in-plane axis between $\Vec{b}_1$ and $\Vec{b}_3$ which requires $(m, n, r, n-r, n', m)$.
The constraints on the zone quantum numbers are proved by a similar argument to the 12-fold case.

In our previous work \cite{oka2021fractal}, we obtained only four zone quantum numbers in hBN/graphene/hBN systems
because we only considered strictly 120$^\circ$-symmetric commensurate approximants.
There the unit areas have relationship $S^*_{24}=S^*_{13}$ and  $S^*_{23}=-S^*_{13}-S^*_{14}$,
so that $\sum_{\langle i,j \rangle} \nu_{ij} S^*_{ij}$ is reduced to $m_1 S^*_{12} + m_2 S^*_{34} + m_3 S^*_{13} + m_4 S^*_{14}$
with $(m_1,m_2,m_3,m_4) = (\nu_{12}, \nu_{34},\nu_{13}-\nu_{23}+\nu_{24}, \nu_{14}-\nu_{23})$,
which are the four integers defined in Ref.\ \onlinecite{oka2021fractal}.
Recalling that $\nu_{ij}$ must have the form $ (m, n, r, n-r, n, m')$ in 120$^\circ$-symmetry,
we can restore the complete six numbers $\nu_{ij}$ as
$(m_1, \frac{2m_3-m_4}{3},\frac{m_3+m_4}{3},\frac{m_3-2m_4}{3},\frac{2m_3-m_4}{3},m_2)$.
The numbers are found to be integers for all the gaps identified in Ref.\ \onlinecite{oka2021fractal}.



\begin{figure*}
\centering
\includegraphics[width=0.9\linewidth]{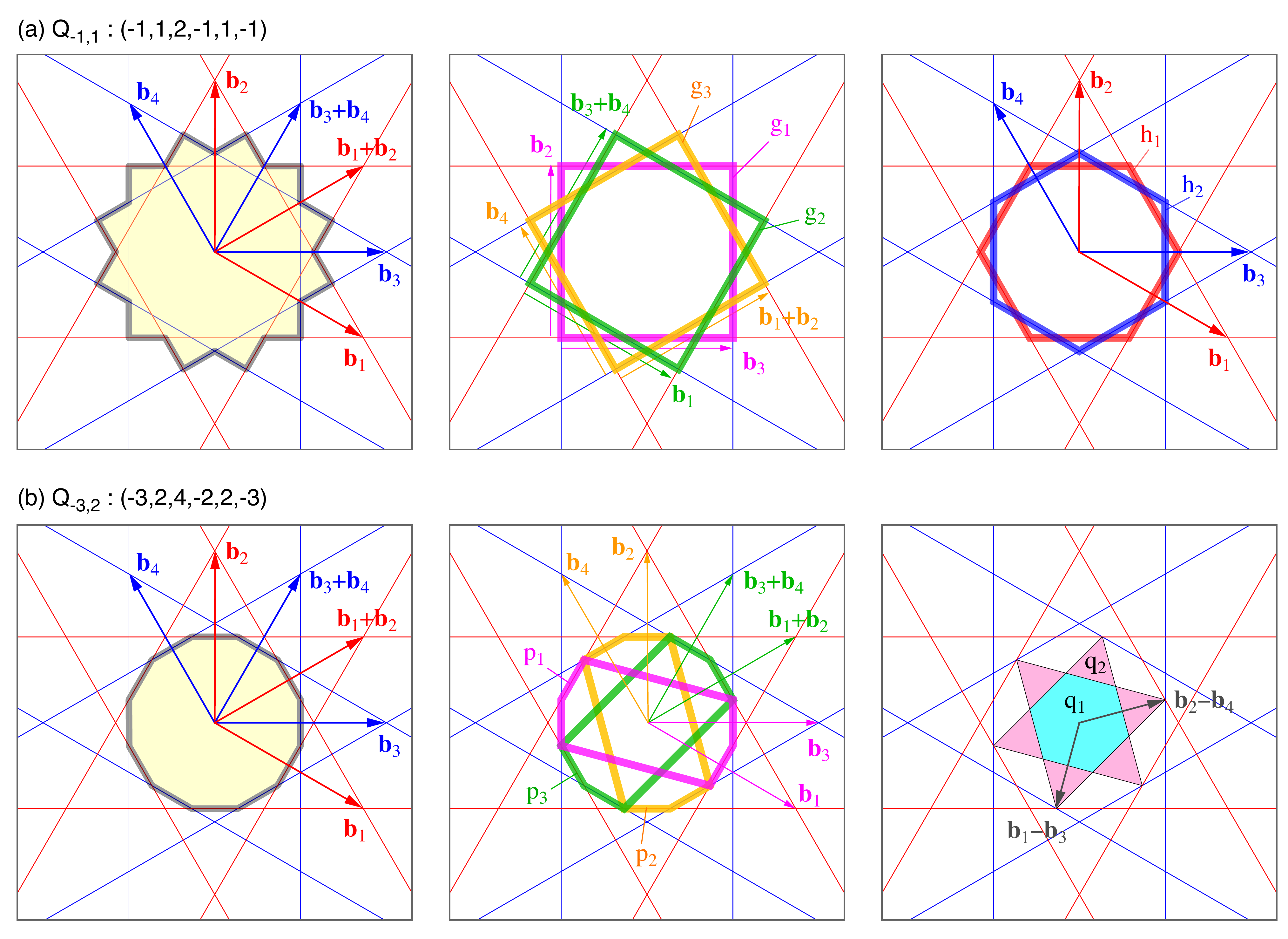}
\caption{The qBZs of (a) $Q_{-1,1}$ and (b) $Q_{-3,2}$ in the twisted triangular potential with $\theta=30^\circ$.
The right two panels in each row illustrate the decomposition of the qBZ into the primitive Brillouin zones.
}
\label{fig_qBZ}
\end{figure*}

\subsection{Quasi Brillouin zones}
\label{sec_qBZ}

The area $\sum_{\langle i,j \rangle} \nu_{ij} S^*_{ij}$ can
be associated with a geometric object in the momentum space referred to as the quasi Brillouin zone (qBZ).
The boundary qBZ for a given gap is defined as a set of $k$-points on the original free-electron band,
at which the gap starts to open in the infinitesimal potential limit. \cite{oka2021fractal}
Generally, the qBZ is a polygon composed of multiple segments of Bragg planes,
which are the perpendicular bisectors of composite reciprocal lattice vectors 
$\Vec{G} = m_1 \Vec{b}_1 + m_2\Vec{b}_2 + m_3 \Vec{b}_3 + m_4 \Vec{b}_4$.

Let us consider the twisted triangular potential considered in the previous section.
The qBZ for the moir\'e gap $M_n$ is found to be just $n$-th Brillouin zone defined by the moir\'e reciprocal vectors
$\Vec{G}^{\rm M}_1$ and $\Vec{G}^{\rm M}_2$ [Eq.\ (\ref{eq_moire_G})].
In general twist angles, however, the qBZ does not coincide with any Brillouin zone of a periodic system.
We show the qBZs of the gap $Q_{-1,1}$ and $Q_{-3,2}$ at $\theta=30^\circ$ in the leftmost panels of Fig. \ref{fig_qBZ}(a) and (b), respectively.
The areas of these qBZs can be easily calculated by the decomposition illustrated in the right two panels.
For instance, the area of the qBZ for gap $Q_{-1,1}$ [Fig. \ref{fig_qBZ}(a)] is expressed by
three squares, $g_1, g_2, g_3$, and two hexagons $h_1, h_2$ as
$S^*(Q_{-1,1}) = (g_1 + g_2 + g_3) - (h_1 + h_2)$.
The area $g_1$ is given by $g_1 = (\Vec{b}_3 \times \Vec{b}_2)_z = - S^*_{23}$,
and similarly we have $g_2 =  S^*_{14} + S^*_{24}$, $g_3 =  S^*_{13} + S^*_{14}$, $h_1 = S^*_{12}$ and $h_2 = S^*_{34}$.
Finally we have 
$S^*(Q_{-1,1})  = - S^*_{12} +S^*_{13} + 2S^*_{14} - S^*_{23} + S^*_{24} -S^*_{34}$,
which agrees with the zone quantum numbers $(-1,1,2,-1,1,-1)$ obtained in the previous section.

Similarly, the area of the qBZ for gap $Q_{-3,2}$ is expressed by
$S^*(Q_{-3,2}) = p_1 + p_2 + p_3 -  2q_1 - q_2$ as shown Fig.\ \ref{fig_qBZ}(b).
The area $p_1$ is the Wigner-Seitz cell in the reciprocal lattice of $\Vec{b}_1$ and $\Vec{b}_3$,
and hence $p_1 = S^*_{13}$.  The $q_1$ (hexagon) and $q_2$ (six triangles)  are the first and second Brillouin zones
defined by the primitive vectors $\Vec{b}_1-\Vec{b}_3$ and $\Vec{b}_2-\Vec{b}_4$,
and therefore $q_1=q_2=[( \Vec{b}_1-\Vec{b}_3)\times(\Vec{b}_2-\Vec{b}_4)]_z=S^*_{12} + S^*_{34} - S^*_{14} + S^*_{23}$.
As a result, the area $S^*(Q_{-3,2}) $ becomes $(-3,2,4,-2,2,-3)$.

At 30$^\circ$, we have the symmetry constraints such as $g_1=g_2=g_3$ and one might think the decomposition of the qBZ area into $S^*_{ij}$'s 
is not unique. However, the area quantization with the same $\nu_{ij}$ strictly holds when the potential is deformed to break the symmetry,
and this guarantees a uniqueness of the decomposition.

\section{Adiabatic charge pumping}
\label{sec_pump}

Here we show that the zone quantum numbers introduced in the previous section
characterize the adiabatic charge pumping under the relative sliding of the doubly periodic potential.

\subsection{1D systems}
\label{sec_pump_1D}

We first consider a doubly-periodic 1D Hamiltonian 
\begin{equation}
H = \frac{p^2}{2m} + V_1(x) + V_2 (x),
\label{eq_H_1D}
\end{equation}
where $V_i(x) = \sum_{m} V_{i,m} \, e^{im b_i x} (i=1,2)$ is a periodic potential with the period of $a_i = 2\pi / b_i$.
Now we consider a cyclic process where one of the periodic potential $V_i(x)$ is adiabatically translated
by its period $a_i$, with the other fixed. 
The translated potential is expressed as 
\begin{equation}
V_i\Bigr (x- \frac{\phi_i}{2\pi}a_i \Bigr) =  \sum_{m} V_{i,m} \, e^{im (b_i x - \phi_i)},
\label{eq_V_1D_slide}
\end{equation}
where an increase of $\phi_i$ from 0 to $2\pi$ gives a unit slide of $V_i(x)$ by distance $a_i$.

We define $\Delta P_i$ by the change of the electric polarization during a unit slide.
In 1D, the $\Delta P_i$ has a dimension of the electronic density (number of electrons per a unit length)
times length, which is dimensionless. Note that we exclude the electric charge $-e$ in the definition of the polarization.
Here we claim the following: When the Fermi energy is in a gap,
the polarization change per cycle, $\Delta P_i$, and the electron density below the gap, $n_e$, are related by
\begin{equation}
\Delta P_i = 2\pi \frac{\partial n_e}{\partial b_i}.
\label{eq_P_vs_ne_1D}
\end{equation}

\begin{figure}
\centering
\includegraphics[width=1.\linewidth]{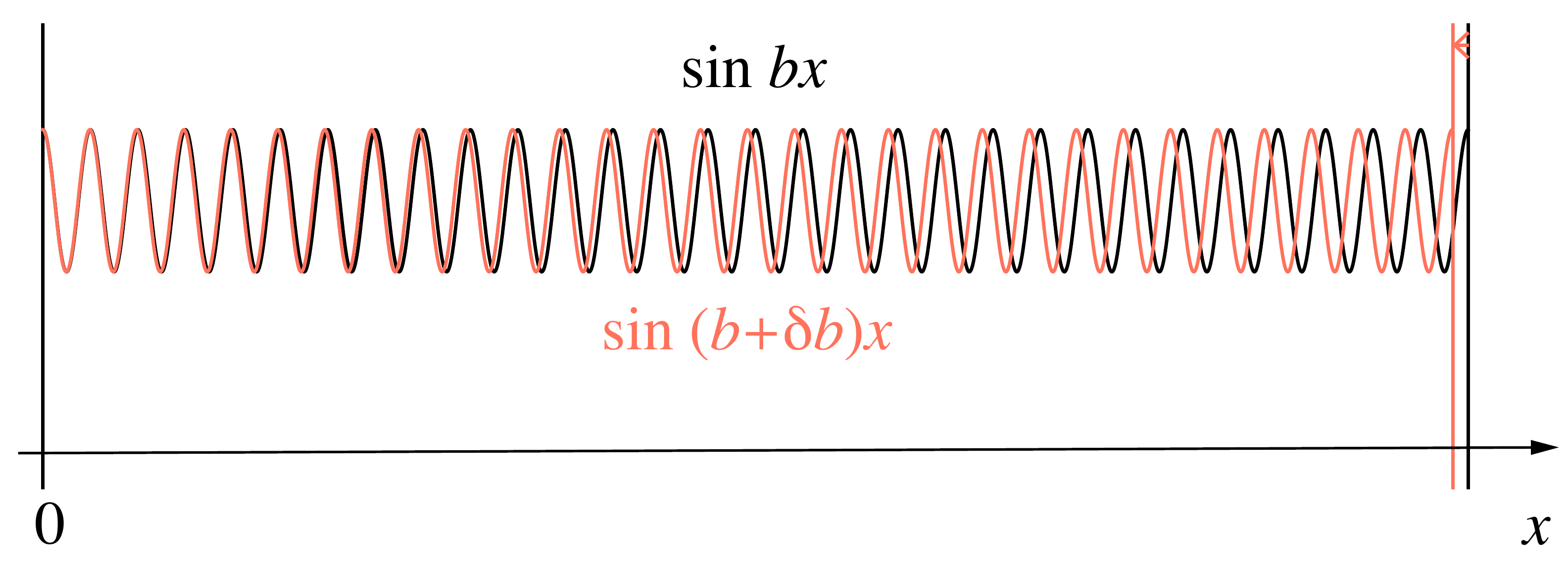}
\caption{
Schamatic picture of an adiabatic process slightly changing the wavenumber of a 1D periodic potential. 
At a point far from the origin ($|x| \gg 2\pi/b$), the change can be viewed as a parallel translation of the potential.
}
\label{fig_schem_slide_1D}
\end{figure}

Eq.\ (\ref{eq_P_vs_ne_1D}) can be proved by the following consideration.
Let us consider an adiabatic process where the wavenumber $b_i$ is slightly changed to $b_i + \delta b_i$.
As illustrated in Fig.\ \ref{fig_schem_slide_1D}, the corresponding change of $V_i(x)$ at a point far from the origin ($|x| \gg a_i$)
can be viewed as a parallel translation of the fixed potential $V_i(x)$.
Considering the phase factor $b_i x - \phi_i$ in Eq.\ (\ref{eq_V_1D_slide}),
the change of  $b_i$ to $b_i + \delta b_i$ can be absorbed to
the change of $\phi_i$ by $\delta \phi_i = - \delta b_i\,x$, and this gives the phase shift in the effective translation at $x$.
Since this process amounts to $n= \delta \phi_i / (2\pi)$ cycles of a unit translation,
the number of electrons passing through the point $x$ is given by
$n \Delta P_i = -\Delta P_i \delta b_i \, x/ (2\pi)$.
Due to the continuity of the electric charge, it must be equal to the change of the number of electrons in a region from 0 to $x$.
This leads to a relation  $-\Delta P_i \delta b_i \, x/ (2\pi) = - x \delta n_e $, 
and we obtain Eq.\ (\ref{eq_P_vs_ne_1D}).

In the doubly-periodic system given by Eq.\ (\ref{eq_H_1D}), 
every gap in the spectrum is characterized by a pair of integers $m_1$ and $m_2$,
such that the electron density below the gap is given by 
\begin{equation}
n_e = \frac{1}{2\pi}(m_1 b_1 + m_2 b_2) = \frac{m_1}{a_1} +  \frac{m_2}{a_2},
\label{eq_ne_1D}
\end{equation}
By using Eq.\ (\ref{eq_P_vs_ne_1D}), 
we conclude $\Delta P_i = m_i$, i.e., $m_i$ electrons passed through any cross section of the system.
The integers $m_1$ and $m_2$ coincide with the first Chern numbers \cite{thouless1983quantization,niu1986quantum,kraus2012topological,fujimoto2020topological},
as presented in Appendix \ref{sec_comm_approx}.

\subsection{2D systems}
\label{sec_pump_2D}

The same argument is available for a doubly-periodic 2D Hamiltonian, Eq.\ (\ref{eq_H_2D}).
We consider an adiabatic translation of the periodic potential $V^\lambda$  ($\lambda= \alpha$ or $\beta$) 
by $\Vec{a}^{\lambda}_{\mu}$,
and calculate the change of the electric polarization during the process. 
A parallel translation of $V^\lambda(\Vec{r})$ is expressed as
\begin{align}
&V^\lambda\Bigl(\Vec{r}-  \frac{\phi^\lambda_1}{2\pi} \Vec{a}^\lambda_1 - \frac{\phi^\lambda_2}{2\pi} \Vec{a}^\lambda_2\Bigr) 
\nonumber\\
& = \sum_{m_1,m_2} 
V^\lambda_{m_1,m_2} \, e^{im_1 (\Vec{b}^\lambda_1\cdot \Vec{r} - \phi^\lambda_1) + im_2 (\Vec{b}^\lambda_2\cdot \Vec{r} - \phi^\lambda_2)}
\label{eq_V_2D_slide}
\end{align}
where we used $\Vec{a}^{\lambda}_{\mu}\cdot \Vec{b}^{\lambda}_{\nu} = 2\pi \delta_{\mu\nu}$.
An increase of $\phi^\lambda_\mu$ from 0 to $2\pi$ gives a unit translation of the potential $V^\lambda$ by $\Vec{a}^{\lambda}_{\mu}$.

The situation can be systematically described by a generalized Hamitonian $H = \Vec{p}^2/(2m) + V$ with
\begin{align}
&V(\Vec{r};\phi_1,\cdots,\phi_N) = \nonumber\\
& \quad \quad \sum_{m_1,\cdots,m_N} V_{m_1,\cdots m_N} \, e^{i\sum_{i=1}^N m_i (\Vec{b}_i \cdot\Vec{r}-\phi_i)}.
\label{eq_V_general}
\end{align}
The current double-period system corresponds to $N=4$, where $\Vec{b}_1, \cdots,\Vec{b}_4$ are given by Eq.\ (\ref{eq_G1234}),
and 
\begin{align}
&(\phi_1,\phi_2,\phi_3,\phi_4) =  (\phi_1^{\alpha}, \phi_2^{\alpha},\phi_1^{\beta}, \phi_2^{\beta}).
\label{eq_s1234}
\end{align}
We consider a cyclic process where $\phi_i$ of a certain $i$ is adiabatically increased from 0 to $2\pi$.
When the Fermi energy is in a gap,
we can show that the change of the electric polarization during the process is
\begin{equation}
\Delta \Vec{P}_i = 2\pi \frac{\partial n_e}{\partial \Vec{b}_i},
\label{eq_P_vs_ne_2D}
\end{equation}
which is the 2D version of Eq.\ (\ref{eq_P_vs_ne_1D}).
Now $\Delta \Vec{P}_i$ has a dimension of the electronic density (number of electrons per a unit area) times length.

Eq.\ (\ref{eq_P_vs_ne_2D}) is derived as follows. 
Let us consider the change of the potential $V(\Vec{r})$ when $\Vec{b}_i$  is changed to $\Vec{b}_i +\delta \Vec{b}_i$.
In a similar manner to 1D case, the change at a point far from the origin ($|\Vec{r}| \gg 2\pi/|\Vec{b}_i|$)
is equivalent to a parallel translation of $\delta \phi_i = - \delta\Vec{b}_i \cdot\Vec{r}$,
noting the phase factor $\Vec{b}_i \cdot\Vec{r}-\phi_i$ in Eq.\ (\ref{eq_V_general}).
This causes a polarization change by $\Delta \Vec{P}_i\delta \phi_i/(2\pi) = \Delta \Vec{P}_i(-\delta\Vec{b}_i \cdot\Vec{r})/(2\pi)$ at the point $\Vec{r}$. 
The number of electrons passing through a line segment from $\Vec{r}$ to  $\Vec{r}+d\Vec{r}$
is given by
\begin{equation}
dN_e =   [(d\Vec{r} \times \Vec{e}_z) \cdot  \Delta \Vec{P}_i] (\delta\Vec{b}_i \cdot\Vec{r})/(2\pi).
\label{eq_dN_e}
\end{equation} 

Now we consider a large closed path $C$ on the 2D plane, and let $N_e$ the number of electrons inside $C$.
When $\Vec{b}_i$ is changed to $\Vec{b}_i+\delta \Vec{b}_i$,
the change of the $N_e$ is calculated by integrating Eq.\ (\ref{eq_dN_e}) along the path, to obtain
\begin{align}
\delta N_e 
& =  \oint_C dN_e = \frac{1}{2\pi} \oint_C  [ (d\Vec{r} \times \Vec{e}_z)\cdot \Delta \Vec{P}_i]  (\delta\Vec{b}_i\cdot\Vec{r})
\nonumber\\
& = \frac{S}{2\pi}  \Delta \Vec{P}_i \cdot \delta\Vec{b}_i 
\end{align}
where $S$ is the area of the region enclosed by $C$
and we used the relationship $\oint_C (d\Vec{r}\times \Vec{e}_z)_\mu  r_\nu = S \delta_{\mu\nu}$
in 2D.
Since $n_e=N_e/S$, we end up with Eq.\ (\ref{eq_P_vs_ne_2D}).
Alternatively, Eq.\ (\ref{eq_P_vs_ne_2D}) can also be derived in the infinitesimal potential limit, 
by integrating the Berry curvature on the boundary of the quasi Brillouin zone.
The detailed argument is presented in Appendix \ref{sec_berry}.

In a 2D doubly-periodic system, the electron density below an energy gap is quantized by Eq.\ (\ref{eq_ne_2D}),
as argued in the previous section.
By applying the formula Eq.\ (\ref{eq_P_vs_ne_2D}) to Eq.\ (\ref{eq_ne_2D}),
the charge pumping $\Delta\Vec{P}_i$ is explicitly calculated as
\begin{align}
\Delta \Vec{P}_i
&= \frac{1}{2\pi}\sum_{\langle k,j \rangle} \nu_{kj} \frac{\partial S^*_{kj}}{\partial \Vec{b}_i} \nonumber\\
&=  \frac{1}{2\pi} \sum_{j} \nu_{ij} (\Vec{b}_j  \times \Vec{e}_z),
\label{eq_P_2D}
\end{align}
where we used $S^*_{ij} =  (\Vec{b}_{i} \times \Vec{b}_{j})\cdot \Vec{e}_z =   (\Vec{b}_{j} \times \Vec{e}_z) \cdot \Vec{b}_{i}$.

By using the real space lattice vectors Eq.\ (\ref{eq_avecs}), Eq.\ (\ref{eq_P_2D}) can also be written as
\begin{align}
\Delta \Vec{P}_i =  \sum_{j}\frac{\nu_{ij}}{S_{ij}}{\Vec{a}^{ij}_1}.
\label{eq_P_2D_2}
\end{align}
The physical interpretation of Eq.\ (\ref{eq_P_2D_2}) is as follows.
Eq.\ (\ref{eq_ne_2D}) states that $\nu_{ij}$ electrons reside in each unit area of $S_{ij}$.
When $\phi_1$ is changed from 0 to $2\pi$ (i.e., $V^\alpha$ is slid by $\Vec{a}^\alpha_1$), for instance,
the wave surface of  $\Vec{b}_1$ is moved by its single period, resulting in shifts of the unit areas  $S_{12}, S_{13}, S_{14}$ 
by $\Vec{a}^{12}_1, \Vec{a}^{13}_1, \Vec{a}^{14}_1$, respectively [See, Fig.\ \ref{fig_unit_areas}(b)].
For each of $j=2,3,4$, the electron density of $\nu_{1j}/S_{1j}$ is transferred by $\Vec{a}^{1j}_1$, 
resulting in a polarization change by $\Delta \Vec{P}_1 = \sum_{j=2,3,4} (\nu_{1j}/S_{1j})\Vec{a}^{1j}_1$.

\subsection{Example: Twisted triangular potentials}

As an example, we consider the adiabatic pumping in the twisted triangular potential in Sec.\ \ref{sec_twisted_trig}.
For the moir\'e gap $M_n = n(1,0,-1,1,0,1)$ [Eq.\ (\ref{eq_M_n})], Eq.\ (\ref{eq_P_2D}) immediately leads to equations 
\begin{align}
&\Delta \Vec{P}_1 =  \frac{n}{2\pi} (\Vec{b}_2-\Vec{b}_4) \times \Vec{e}_z = \frac{n}{S_{\rm M}}  \Vec{L}^{\rm M}_1,\nonumber\\
&\Delta \Vec{P}_2 =  \frac{n}{2\pi} (\Vec{b}_1-\Vec{b}_3) \times \Vec{e}_z = \frac{n}{S_{\rm M}}  \Vec{L}^{\rm M}_2,\nonumber\\
&\Delta \Vec{P}_3 =  -\frac{n}{2\pi} (\Vec{b}_2-\Vec{b}_4) \times \Vec{e}_z = -\frac{n}{S_{\rm M}}  \Vec{L}^{\rm M}_1,\nonumber\\
&\Delta \Vec{P}_4 =  -\frac{n}{2\pi} (\Vec{b}_1-\Vec{b}_3) \times \Vec{e}_z = -\frac{n}{S_{\rm M}}  \Vec{L}^{\rm M}_2,
\label{eq_pump_M_n}
\end{align}
where $\Vec{L}^{\rm M}_i$ is the moir\'e lattice vector defined by
\begin{align}
\Vec{L}^{\rm M}_1  = \frac{S_{\rm M}}{2\pi} (\Vec{G}^{\rm M}_2 \times \Vec{e}_z),
\,\,
\Vec{L}^{\rm M}_2  = - \frac{S_{\rm M}}{2\pi}  (\Vec{G}^{\rm M}_1 \times \Vec{e}_z),
\end{align}
and we used Eq.\ (\ref{eq_moire_G}).
This 
indicates that,
when the potential $\alpha(\beta)$ is slid by its unit vector $\Vec{a}^\alpha_\mu(\Vec{a}^\beta_\mu)$,
then $n$ electrons per the moir\'e unit cell are pumped by a moir\'e unit vector $n\Vec{L}^{\rm M}_\mu(-n\Vec{L}^{\rm M}_\mu)$.
The result coincides with the adiabatic moir\'e  pumping in the previous works.\cite{fujimoto2020topological,zhang2020topological,su2020topological}

The argument is also applicable to the quasicrystal gaps at $\theta=30^\circ$.
Here the zone quantum numbers take the form  $Q_{m,n} =  (m, n, 2n, -n, n, m)$ [Eq.\ (\ref{eq_Q_ab})].
For $\Delta \Vec{P}_1$, for instance, Eq.\ (\ref{eq_P_2D}) gives
\begin{align}
\Delta \Vec{P}_1 &=  \frac{1}{2\pi} (m\Vec{b}_2 + n\Vec{b}_3 + 2n\Vec{b}_4) \times \Vec{e}_z \nonumber\\
&= \frac{1}{L} (m+\sqrt{3} n) 
\begin{pmatrix}
1 \\ 0
\end{pmatrix}
=
\frac{n_e}{2}  \Vec{a}^{\alpha}_1.
\label{eq_P1_30}
\end{align}
In the last equation, we used Eq.\ (\ref{eq_trig_L}) and note that the electronic density [Eq.\ (\ref{eq_ne_2D})]
is $n_e = (\sqrt{3}m+3n)/L^2$ at $\theta=30^\circ$.
By similar calculations, we have a set of equations independent of $m$ and $n$,
\begin{align}
&\Delta \Vec{P}_1 =  \frac{n_e}{2}  \Vec{a}^{\alpha}_1,\quad
\Delta \Vec{P}_2 =  \frac{n_e}{2}  \Vec{a}^{\alpha}_2,\nonumber\\
&\Delta \Vec{P}_3 =  \frac{n_e}{2}  \Vec{a}^{\beta}_1,\quad
\Delta \Vec{P}_4 =  \frac{n_e}{2}  \Vec{a}^{\beta}_2.
\label{eq_pump_Q_ab}
\end{align}

From the definition, $\Delta \Vec{P}_1$ and $\Delta \Vec{P}_2$ ($\Delta \Vec{P}_3$ and $\Delta \Vec{P}_4$) represent
the polarization changes when the potential $\alpha$ ($\beta$) is translated by $\Vec{a}^{\alpha}_1$ and $\Vec{a}^{\alpha}_2$
($\Vec{a}^{\beta}_1$ and $\Vec{a}^{\beta}_2$), respectively.
Eq.\ (\ref{eq_pump_Q_ab}) shows that, in any sliding processes, the transfer of the electric charge is always parallel to the potential sliding direction
(regardless of which potential we move),
and that the amount of the charge pumping is equivalent
to the movement of the half of the total electric charge by the sliding vector.

\section{4D quantum Hall effect and the second Chern numbers}
\label{sec_4D}

In the following, we describe the adiabatic pumping argued in the previous section in an alternative approach
using the dimensional reduction of the four-dimensional (4D) quantum Hall effect (QHE), \cite{kraus2013four,qi2008topological,price2015four,price2016measurement,lohse2018exploring,zilberberg2018photonic,lu2018topological},
and demonsrate that the zone quantum number $\nu_{ij}$ coincides with the second Chern number.

\begin{figure}
\centering
\includegraphics[width=0.8\linewidth]{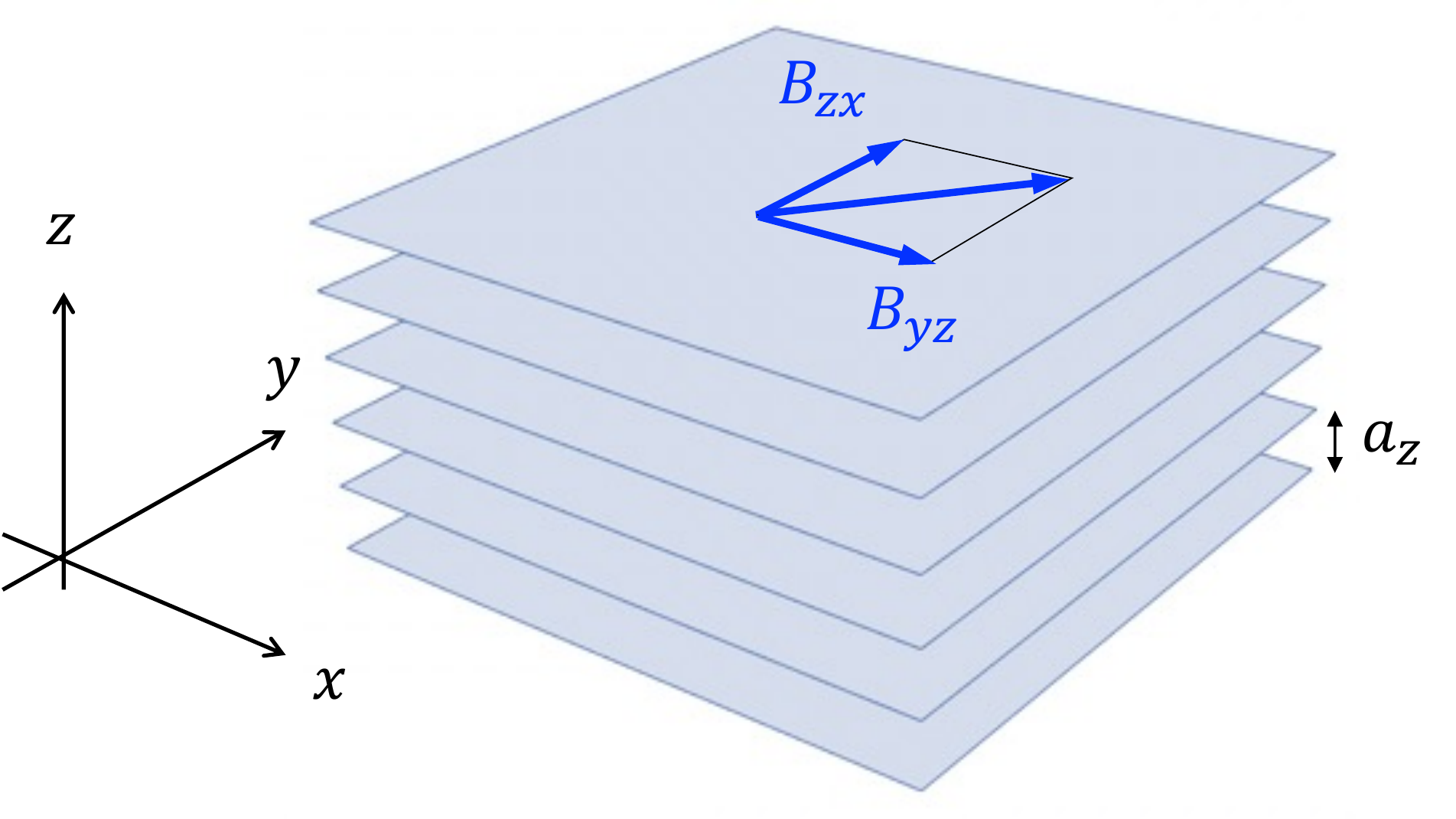}
\caption{An infinite stack of 2D free-electron systems under in-plane magnetic field $\Vec{B} = (B_{yz}, B_{zx}, 0)$.
}
\label{fig_3d_qhe}
\end{figure}

We first consider the 3D QHE as a simple example.
Let us consider an infinite stack of 2D free-electron systems as illustrated Fig.\ \ref{fig_3d_qhe},
which is continuous in $x$ and $y$ directions and discrete in $z$ direction with lattice spacing $a_z$.
For $z$-direction, we assume the nearest-neighbor tight-binding coupling $t_z$ between the adjacent layers.
We apply a magnetic field $B_{\mu\nu} = \partial_\mu A_\nu - \partial_\nu A_\mu$.
Here we assume a uniform, in-plane field $\Vec{B} = (B_{yz}, B_{zx}, 0)$,
and set the vector potential as $\Vec{A}=(0,0,A_z)$ with $A_z = B_{xz} x + B_{yz} y$ (note $B_{xz}=-B_{zx}$).
The motion of an electron is described by the Schr\"{o}dinger equation,
\begin{align}
&\frac{\Vec{p}^2}{2m}\Psi (x,y,z) 
- t_z
\Bigl[
e^{i\frac{e}{\hbar}A_z a_z} \Psi(x,y,z+a_z)  
\nonumber\\
&\quad  + e^{-i\frac{e}{\hbar}A_z a_z} \Psi(x,y,z-a_z) 
\Bigr] = E  \Psi (x,y,z),
\label{eq_schr_3D_layer}
\end{align}
where $\Vec{p} = -i\hbar(\partial_x, \partial_y)$ is the in-plane momentum.
Since the Hamiltonian is periodic in $z$, the wavefunction can be factorized as
$\Psi (x,y,z) = \psi(x,y) e^{ik_z z}$,
and then Eq.\ (\ref{eq_schr_3D_layer}) is reduced to a 2D Schr\"{o}dinger equation,
\begin{align}
&\frac{\Vec{p}^2}{2m} \psi - 2t_z \cos(\Vec{b}\cdot \Vec{x} + \phi_z) \psi = E  \psi,
\label{eq_schr_3D_layer_reduced}
\end{align}
where  $\Vec{b} = (ea_z/\hbar) (B_{xz},B_{yz})$, $\Vec{x}=(x,y)$ and $\phi_z = k_z a_z$.
This is a 2D system with a single sinusoidal potential with the wave number $\Vec{b}$.
The phase factor $\phi_z$ corresponds to the wavenumber in $z$ direction.

The extension of higher dimensions is straightforward.
We consider a six-dimenisonal (6D) system in $(x, y, z_1,z_2,z_3,z_4)$ space,
which is continuous in $x$ and $y$ directions and discrete in $z_i\,(i=1,2,3,4)$ direction.
We apply a uniform magnetic field $B_{yi}$ and $B_{ix}(=-B_{xi})$ on $y z_i$-plane and $z_i x$-plane, respectively,
and take the vector potential $\Vec{A} = \sum_{i=1}^4 ( B_{xi} x + B_{yi} y) \Vec{e}_i$ where $\Vec{e}_i$ is the unit vector in $z_i$ direction.
As the Hamiltonian is periodic in any $z_i$'s, the wavefunction can be written as
$\Psi (x,y,z_1,z_2,z_3,z_4) = \psi(x,y) e^{i\sum_i k_i z_i}$,
where $k_i$ is the Bloch wavenumber defined in $\pi/a_i < k_i \leq \pi/a_i$. 
The 6D Schr\"{o}dinger equation is reduced to $(x,y)$ space  as
\begin{align}
&\frac{\Vec{p}^2}{2m} \psi - \sum_{i=1}^4 2t_i \cos(\Vec{b}_i\cdot \Vec{x} + \phi_i) \psi = E  \psi,
\label{eq_schr_3D_layer_reduced_4D}
\end{align}
where
\begin{align}
\Vec{b}_i = \frac{e  a_i}{\hbar} (B_{xi},B_{yi}), \quad \phi_i = k_i a_i.
\label{eq_b_and_B}
\end{align}
This is equivalent to the double-period 2D system considered in this paper.
The higher harmonic terms in $\Vec{b}_i$ can be incorporated by assuming the further layer hopping in $z_i$ direction.

The electromagnetic response of the system is characterized by the second Chern number. \cite{kraus2013four,qi2008topological,price2015four,price2016measurement,lohse2018exploring,zilberberg2018photonic,lu2018topological} 
Let us consider a commensurate approximant where the periodicities of $\Vec{b}_i (i=1,2,3,4)$ have a common super unit cell,
and define the Bloch wavenumber $(k_x,k_y)$ in the corresponding super Brillouin zone.
The Bloch Hamiltonian for the 6D system is written as $H(k_x,k_y,k_1,k_2,k_3,k_4)$.
We consider the 4D subspace $k_\mu=(k_x,k_y,k_i,k_j)$ by choosing two indexes $i,j$ from 1 to 4,
with the rest two wavenumbers fixed. 
When the spectrum of 4D Hamiltonian $H(k_x,k_y,k_i,k_j)$ is gapped, 
the second Chern number for the gap is defined as \cite{kraus2013four,qi2008topological,price2015four,price2016measurement,lohse2018exploring,zilberberg2018photonic,lu2018topological},
\begin{align}
C_{ij}^{(2)} = \frac{1}{32\pi^2}
\int_{\rm BZ} d^4k \, \epsilon^{\mu\nu\lambda\rho} 
{\rm Tr}
[{\cal F}_{\mu\nu} {\cal F}_{\lambda\rho}] \,\,\in \mathbb{Z}.
\label{eq_C2}
\end{align}
Here ${\rm BZ}$ stands for the 4D Brillouin zone (a 4D torus), $\epsilon^{\mu\nu\lambda\rho}$ is the antisymmetric tensor of rank 4 and 
${\cal F}_{\mu\nu}$ is a matrix defined by
\begin{align}
&{\cal F}_{\mu\nu}^{\alpha\beta} 
= \partial_\mu {\cal A}_{\nu}^{\alpha\beta} - \partial_\nu {\cal A}_{\mu}^{\alpha\beta} + i [{\cal A}_{\mu}, {\cal A}_{\nu}]^{\alpha\beta},
\nonumber\\
&{\cal A}_{\mu}^{\alpha\beta}(\Vec{k})  = -i \langle \alpha, \Vec{k} | \partial_\mu |  \beta, \Vec{k} \rangle,
\end{align}
where $\partial_\mu = \partial/\partial k_\mu$, $|\alpha, \Vec{k} \rangle$ is the eigenstates of the $\alpha$-th band,
and the indeces $\alpha$ and $\beta$ run over all the bands below the gap.
 It is alternatively expressed as \cite{kraus2013four}
 \begin{align}
C_{ij}^{(2)} = -\frac{1}{8\pi^2}
\int_{\rm BZ} d^4k \, \epsilon^{\mu\nu\lambda\rho} 
{\rm Tr}\Bigl[
P \frac{\partial P}{\partial k_\mu}\frac{\partial P}{\partial k_\nu}
P \frac{\partial P}{\partial k_\lambda}\frac{\partial P}{\partial k_\rho}
\Bigr]
\label{eq_C2-2}
\end{align}
where $P(\Vec{k}) = \sum_{\alpha \in {\rm occ}} |  \alpha, \Vec{k} \rangle \langle \alpha, \Vec{k} |$ 
is the projection operator to the eigenstates below the gap.
Note that we have six second Chern numbers 
depending on the choice of $i,j (i \neq j)$ from 1,2,3,4.

When the Fermi energy is in the gap, the electro-magnetic response of the 4D system is given by 
\cite{kraus2013four,qi2008topological,price2015four,price2016measurement,lohse2018exploring,zilberberg2018photonic,lu2018topological} 
\begin{align}
j_\mu^{\rm (4D)} = \frac{e^3}{h^2} C_{ij}^{(2)} \epsilon^{\mu\nu\lambda\rho} B_{\nu\lambda} E_{\delta},
\label{eq_j_4D}
\end{align}
where $j_\mu^{\rm (4D)}$ is the electric current density in the 4D space.
If a weak electric field $E_i$ is applied to the system,
the wavenumber $k_i$ is adiabatically changed to $k_i + (e/\hbar) A_i(t)$, where $E_i = -\partial A_i / \partial t$.
When we consider a cyclic process where $\phi_i=k_i a_i$ is changed from 0 to $2\pi$ in a time period $T$,
the corresponding electric field should be  
\begin{align}
E_i = - \frac{h}{e a_i} \frac{1}{T}.
\label{eq_E_j}
\end{align}
According to  Eq.\ (\ref{eq_j_4D}), $E_i$ induces an electric current
$(j_x, j_y)^{\rm (4D)} = (e^3/h^2) C_{ij}^{(2)} (-B_{y j}, B_{x j})E_{i}$.
The corresponding 2D current density per a single layer is given by $j_\mu^{\rm (2D)} = j_\mu^{\rm (4D)} a_ia_j$, giving
\begin{align}
(j_x, j_y)^{\rm (2D)} = -\frac{e}{2\pi} C_{ij}^{(2)} (b_{j,y}, -b_{j,x})\frac{1}{T},
\end{align}
where we used Eqs.\ (\ref{eq_b_and_B}) and (\ref{eq_E_j}).
Total polarization change in the process ($\phi_i : 0 \to 2\pi$) is $\Delta \Vec{P}_i = \Vec{j}^{\rm (2D)}T/(-e)$.
By appending the summation over $j$, we finally obtain
\begin{align}
\Delta \Vec{P}_i =  -\frac{1}{2\pi} \sum_{j} C^{(2)}_{ij} (\Vec{b}_j  \times \Vec{e}_z).
\label{eq_P_2D_Chern}
\end{align}
We notice that Eq.\ (\ref{eq_P_2D_Chern}) has the exactly the same form as Eq.\ (\ref{eq_P_2D}).
By comparing the two equations, we immediately find
\begin{align}
\nu_{ij} = -C^{(2)}_{ij},
\end{align}
i.e., the zone quantum numbers turned out to be the second Chern numbers.

By using Eq.\ (\ref{eq_C2-2}), we numerically calculated $C^{(2)}_{ij}$ 
for some of commensurate approximants in the twisted triangular potential series considered 
in Sec.\ \ref{sec_twisted_trig}
, and confirmed the agreement with $-\nu_{ij}$.
Since Eq.\ (\ref{eq_C2-2}) includes the integral on the Brillouin zone of the commensurate approximant,
one might think that $C^{(2)}_{ij}$ explicitly depends on the Brillouin zone size (inverse of the commensurate unit cell size), which is rather arbitrary
as seen in Appendix \ref{sec_comm_approx}.
But in reality, the integrand ${\rm Tr}[\cdots]$ itself is proportional to the number of the subbands below the gap (proportional to the unit cell size), 
and this cancels with the Brillouin zone integral, giving the invariant integers independent of the commensurate period.



\section{Conclusion}
\label{sec_concl}

We have shown that energy gaps in two-dimensional double-periodic systems can be uniquely labelled by six second Chern numbers.
Physically, these numbers can be interpreted as zone quantum numbers,
which quantize the momentum space in units of the six fundamental Brillouin zones defined in the redundant periodicities.
At the same time, the zone quantum numbers also describe the quantized charge pumping under a relative slide of different periodic potentials.
By considering a mapping of the 2D charge pumping to the fictitious 4D quantum Hall effect,
we found the zone quantum numbers are equivalent to the second Chern numbers.

The topological characterization of energy gaps presented in this work
is applicable to any quasiperiodic systems having redundant reciprocal vectors more than the spatial dimension.
In a twisted multilayer system composed on $n$ layers, for instance,
$2n$ reciprocal vectors define $n(2n-1)$ independent Brillouin zones, so that there should be $n(2n-1)$ zone quantum numbers. 
The Penrose tile \cite{penrose1974role, walter2009crystallography} has 5 reciprocal lattice vectors, giving the 10 quantum numbers.
The extension to 3D quasicrystal should also be possible. 
Lastly, non-zero quantum numbers for adiabatic pumping generally implies the existence of the edge localized states.\cite{hatsugai2016bulk,fujimoto2021moire,kraus2012topological,kraus2013four}
The study of the edge states in general quasiperiodic systems in terms of the zone quantum numbers would also be intriguing.

\appendix

\section{Commensurate approximant method}
\label{sec_comm_approx}

We describe the commensurate approximant method to calculate the band 
structures and the zone quantum numbers in the double-period system.
In an incommensurate case, we always have lattice points of the two periodic potentials
which happen to be very close to each other. The situation is expressed as
\begin{equation}
p_1 \Vec{a}^{\alpha}_1 + p_2 \Vec{a}^{\alpha}_2 = p_3 \Vec{a}^{\beta}_1 + p_4 \Vec{a}^{\beta}_2 + \Delta \Vec{L},
\label{eq_nearly_commensurate}
\end{equation}
where $p_i \,\, (i=1,2,3,4)$ are integers and \(\Delta\Vec{L}\) is the difference.

A commensurate approximant can be obtained by 
choosing two such nearly-commensurate points [with integers $(p_1,p_2,p_3,p_4)$ and $(q_1,q_2,q_3,q_4)$],
and deforming the potential $\beta$ such that  \(\Delta\Vec{L}\) becomes zero.
The two points then become the exact primitive lattice vectors of the commensurate approximant,
\begin{eqnarray}
\left(
\begin{array}{c}
\Vec{a}^{\rm c}_1 \\ \Vec{a}^{\rm c}_2
\end{array}
\right)
&=&
\left( \begin{array}{cc}
p_1 & p_2 \\ 
q_1 & q_2
\end{array} \right)
\left( \begin{array}{c}
\Vec{a}^{\alpha}_1 \\ \Vec{a}^{\alpha}_2
\end{array} \right)
\nonumber\\
&=& \left( \begin{array}{cc}
p_3 & p_4\\
q_3 & q_4
\end{array} \right)
\left( \begin{array}{c}
\Vec{a}^{\beta}_1 \\ \Vec{a}^{\beta}_2
\end{array} \right).
\end{eqnarray}
Correspondingly, the reciprocal superlattice vectors 
\(\Vec{b}^{\rm c}_1, \Vec{b}^{\rm c}_2\) are given by
\begin{eqnarray}
\left(
\begin{array}{c}
\Vec{b}^{\rm c}_1 \\ \Vec{b}^{\rm c}_2
\end{array}
\right)
&=&
\left[
\left( \begin{array}{cc}
p_1 & p_2 \\ 
q_1 & q_2
\end{array} \right)^T
\right]^{-1}
\left( \begin{array}{c}
\Vec{b}^{\alpha}_1 \\ \Vec{b}^{\alpha}_2
\end{array} \right)
\nonumber\\
&=&
\left[
\left( \begin{array}{cc}
p_3 & p_4\\
q_3 & q_4
\end{array} \right)^T
\right]^{-1}
\left( \begin{array}{c}
\Vec{b}^{\beta}_1 \\ \Vec{b}^{\beta}_2
\end{array} \right),
\label{eq_G_and_G_C}
\end{eqnarray}
where $T$ stands for the matrix transpose.
$\Vec{a}^{\rm c}_\mu$  and $\Vec{b}^{\rm c}_\mu$ are related by
\begin{align}
\Vec{a}^{\rm c}_1 = \frac{S_{\rm c}}{2\pi} (\Vec{b}^{\rm c}_2 \times \Vec{e}_z),\quad
\Vec{a}^{\rm c}_2 = -\frac{S_{\rm c}}{2\pi} (\Vec{b}^{\rm c}_1 \times \Vec{e}_z),
\label{eq_L_C_and_G_C}
\end{align}
where $S_{\rm c}= (\Vec{a}^{\rm c}_1 \times \Vec{a}^{\rm c}_2)_z$ is the unit area of the commensurate approximant.

By using the serial notation Eq.\ (\ref{eq_G1234}), Eq.\ (\ref{eq_G_and_G_C}) can simply be written as 
\begin{equation}
\Vec{b}_i = p_i \Vec{b}^{\rm c}_1 + q_i \Vec{b}^{\rm c}_2.
\label{eq_G_C}
\end{equation}
Accordingly, the unit areas Eq.\ (\ref{eq_S_star}) become
\begin{equation}
S^*_{ij} = (p_i q_j - p_j q_i) S^*_{\rm c},
\end{equation}
where $S^*_{\rm c} = (\Vec{b}^{\rm c}_1 \times \Vec{b}^{\rm c}_2)_z = (2\pi)^2/S_{\rm c}$ is the area of the first Brillouin zone of the 
commensurate approximant.
Eq.\ (\ref{eq_ne_2D}) becomes the Diophantine equation,
 \begin{equation}
r = \sum_{\langle i,j \rangle} \nu_{ij} (p_i q_j - p_j q_i),
\label{eq_diophantine}
\end{equation}
where $r \equiv n_e/[S^*_{\rm c}/(2\pi)^2]$ is for the number of the bands below the gap.

\begin{figure*}
\centering
\includegraphics[width=1.\linewidth]{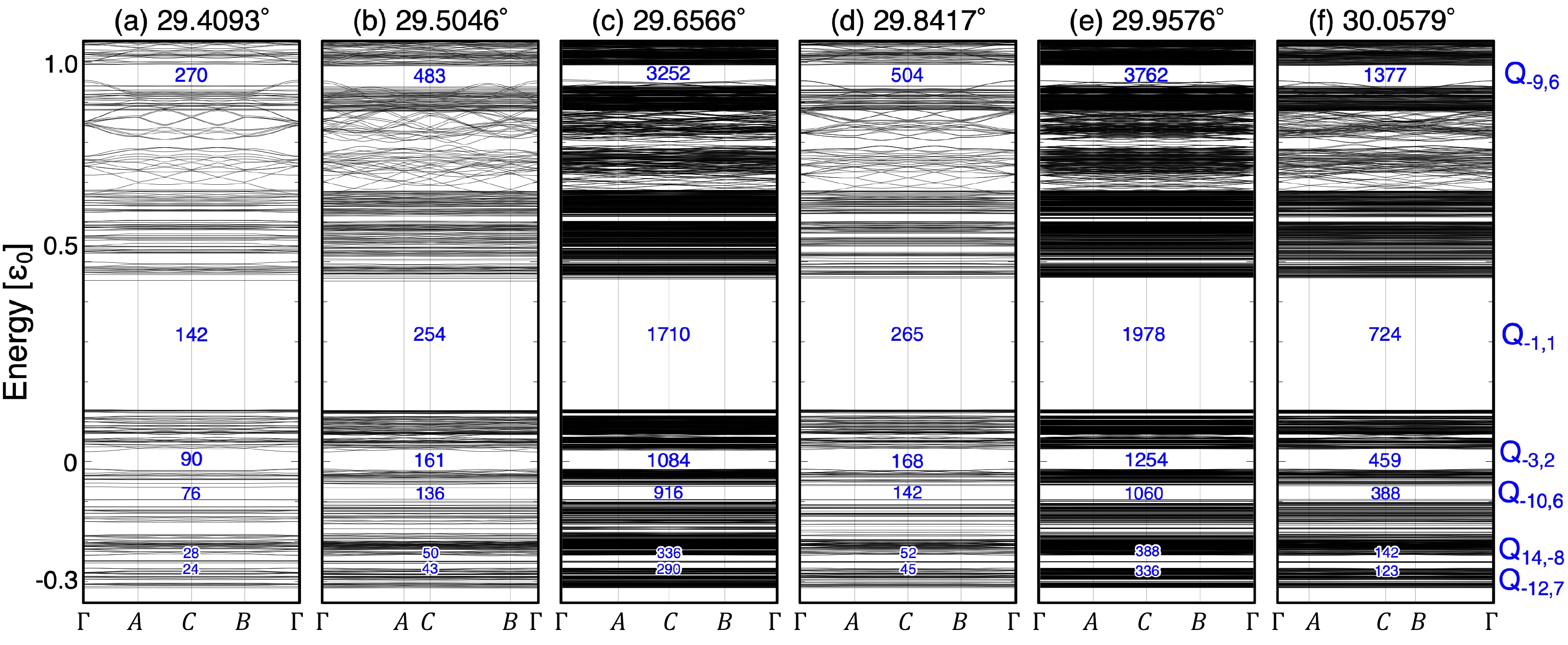}
\caption{
Band structures of commensurate approximants (a) to (f) (specified in Table \ref{tbl_pq}) for the twisted double triangular potential
near 30$^\circ$ degree.
The Brillouin zone path is taken as $(\Gamma,A,C,B,\Gamma) \equiv (0, \Vec{b}^{\rm c}_1/2, (\Vec{b}^{\rm c}_1+\Vec{b}^{\rm c}_2)/2, \Vec{b}^{\rm c}_2/2, 0)$.
The integers in gaps indicate the number of the bands below the gap, $r$.
}
\label{fig_ptri_band}
\end{figure*}

In determination of the zone quantum numbers $\nu_{ij}$, we consider a series of commensurate approximants near the target system, and solve a set of Diophantine equations Eq.\ (\ref{eq_diophantine}) for all the approximants. 
As an example, we show in Fig.\ \ref{fig_ptri_band} the band structures of six commensurate approximants (a) to (f) for the double triangular potential near 
$\theta=30^\circ$ [see, Fig.\ \ref{fig_spectrum_triangular}], which are specified by $(p_1,p_2,p_3,p_4; q_1,q_2,q_3,q_4)$ in Table \ref{tbl_pq}.
The Brillouin zone path is taken as $(\Gamma,A,C,B,\Gamma) \equiv (0, \Vec{b}^{\rm c}_1/2, (\Vec{b}^{\rm c}_1+\Vec{b}^{\rm c}_2)/2, \Vec{b}^{\rm c}_2/2, 0)$.
Table \ref{tbl_pq} also shows the number of the occupied bands $r$ for some major gaps $Q_{m,n}$.
The six systems have very close potential profiles and similar spectral structures, 
while it have completely different sizes of the commensurate unit cells and thus different numbers of bands below the same gap.
For the largest gap $Q_{-1,1}$, for instance, the number of the bands are $r=142, 254, 1710, 265, 1978, 724$ for the six systems,
and accordingly we have six independent equations of Eq.\ (\ref{eq_diophantine}) with six unknown variables $\nu_{ij}$.
By solving the set of the equations, we find $\nu_{ij} = (-1,1,2,-1,1,-1)$ as a unique solution.
All other approximants sharing the same gap have the same solution of $\nu_{ij}$.

\begin{table*}
$
\begin{array}{c|c|cccccccc|cccccc}
& \theta & p_1 & p_2 & p_3 & p_4 & q_1 & q_2 & q_3 & q_4 & r[Q_{-12,7}] & r[Q_{14,-8}] & r[Q_{-10,6}] & r[Q_{-3,2}] & r[Q_{-1,1}]  & r[Q_{-9,6}] \\
\hline
{\rm (a)} & 29.4093 & 3 & 8    & 8 & 3 & -8    & 11 & -3 & 11 & 24 & 28 & 76 & 90 & 142 & 270 \\  
{\rm (b)} & 29.5046 & 3 & 8    & 8 & 3 & -9    & 34 & 9 & 25 & 43 & 50 & 136 & 161 & 254 & 483  \\
{\rm (c)} & 29.6566 & 25 & 9    & 34 & -9 & -27 & 37    & -10 & 37 & 290 & 336 & 916 & 1084 & 1710 & 3252 \\
{\rm (d)} & 29.8417 & 11 & 4    & 15 & -4 & -4 & 15    & 4 & 11 & 45 & 52 & 142 & 168 & 265 & 504 \\
{\rm (e)} & 29.9576 & 11 & 30    & 30 & 11& -30 & 41     & -11 & 41 & 336 & 388 & 1060 & 1254 & 1978 & 3762 \\
{\rm (f)} & 30.0579 & 11 & 30    & 30 & 11& -11 & 15     & -4 & 15 & 123 & 142 & 388 &459 & 724 & 1377 
\end{array}
$
\caption{Twist angle $\theta$ and the indeces $(p_1,p_2,p_3,p_4; q_1,q_2,q_3,q_4)$
of the commensurate approximants (a) to (f).
The $r[Q_{m,n}]$ is the number of the occupied bands below the gap $Q_{m,n}$.
}
\label{tbl_pq}
\end{table*}

The formula of quantum pumping Eq.\ (\ref{eq_P_2D}) can also be transformed to the commensurate version.
By using Eq.\ (\ref{eq_G_C}), Eq.\ (\ref{eq_P_2D}) is written as
\begin{align}
\Delta \Vec{P}_i
&=  \frac{1}{2\pi} \sum_{j(\neq i)} \nu_{ij} 
\bigl[p_j (\Vec{b}^{\rm c}_1\times \Vec{e}_z)  +
 q_j (\Vec{b}^{\rm c}_2\times \Vec{e}_z)\bigr].
\end{align}
By using Eq.\ (\ref{eq_L_C_and_G_C}), it is reduced to
\begin{align}
\Delta \Vec{P}_i
=  \frac{1}{S_{\rm c}} (C_{i1}\Vec{a}^{\rm c}_1  + C_{i2} \Vec{a}^{\rm c}_2),
\label{eq_pump_comm}
\end{align}
where
\begin{align}
C_{i1} =  \sum_{j(\neq i)} \nu_{ij} q_j, \quad 
C_{i2} =  - \sum_{j(\neq i)} \nu_{ij} p_j
\label{eq_C_ij}
\end{align}
are the integers to characterize the pumping in units of the commensurate period.
By using Eqs.\ (\ref{eq_C_ij}) and (\ref{eq_diophantine}), we have the Diophantine equation for $C_{il}$'s,
\begin{align}
&\sum_{i=1}^4 p_i C_{i1} =  r,\quad \sum_{i=1}^4 p_i C_{i2} =  0, \nonumber\\ 
&\sum_{i=1}^4 q_i C_{i1} =  0,\quad \sum_{i=1}^4 q_i C_{i2} =  -r,
\end{align}
which agrees with the results in the previous work. \cite{fujimoto2020topological} 

The integers $C_{i1}$ and $C_{i2}$ are expressed as the first Chern numbers.\cite{fujimoto2020topological,zhang2020topological,su2020topological}
The Bloch Hamiltonian for the commensurate approximant can be written as
$H(k_1, k_2; \phi_1, \phi_2,\phi_3,\phi_4)$ where 
$k_l = \Vec{k}\cdot \Vec{a}^c_l/|\Vec{a}^c_l|\,(l=1,2)$ is the component of the Bloch wavevector along $\Vec{a}^c_l$,
and $\phi_i \, (i=1,2,3,4)$ is the phase factors for the potential slide [Eq.\ (\ref{eq_s1234})].
Then $C_{il}$ is given by the first Chern number on a 2D torus of $(k_l, \phi_i)$, or
\begin{equation}
C_{il} = \frac{1}{2\pi} 
\int_0^{|\Vec{b}^c_l|} dk_l \int_0^{2\pi} d\phi_i
 \,
F_{il} \,\, \in \mathbb{Z},
\label{eq_first_chern}
\end{equation}
where $F_{il}$ is the Berry curvature defined by
\begin{align}
&F_{il} = \partial_1 a^{(2)} - \partial_2 a^{(1)}, \nonumber\\
&a^{(\mu)} =-i \sum_{n\in {\rm occ}}  \langle \alpha,\Vec{k} | \partial_\mu | \alpha, \Vec{k} \rangle,
\end{align}
and $\partial_1 = \partial/\partial k_l$ and $\partial_2 = \partial/\partial \phi_i$.
The integral period of $k$ $(0 \leq k \leq |\Vec{b}^c_l|)$ in Eq.\ (\ref{eq_first_chern})
represents the span of the first Brillouin zone in $l$ direction.

Unlike the second Chern number $\nu_{ij}$,
the first Chern umber $C_{il}$ of the 2D commensurate system directly depends on the unit cell size
and it is not an invariant in a continuous deformation.
In Eq.\ (\ref{eq_C_ij}), indeed, $C_{ij}$ depends on $p_i$ and $q_i$,
and hence the systems in Fig.\ \ref{fig_ptri_band} have all different $C_{ij}$'s for the same gap.
The direct dependence of $C_{ij}$  on the unit cell size
can be understood by considering the same system with a redundant unit cell spanned by
$M_1 \Vec{a}^{\rm c}_1$ and $M_2 \Vec{a}^{\rm c}_2$ with arbitrary integers $M_1$ and $M_2$.
Due to the band folding, the integral path in Eq.\ (\ref{eq_first_chern}) is reduced to $0 \leq k_l \leq |\Vec{b}^c_l|/M_l$,
and the integrand $F_{il}$ (proportional to the number of bands) is multiplied by $M_1M_2$.
As a consequence, the first Chern number for the enlarged unit cell becomes  $C'_{i1} = M_2 C_{i1}$ and $C'_{i2} = M_1 C_{i2}$.
This is natural because the integer $C_{i1}(C_{i2})$ corresponds to the number of electrons passing 
through the unit-cell side along $\Vec{a}^{\rm c}_2(\Vec{a}^{\rm c}_1)$ during a cyclic process,
and hence it is just proportional to the span of the corresponding unit cell side.
In contrast, the second Chern number $\nu_{ij}$ [Eq.\ (\ref{eq_C2})] includes an integral on whole the 2D Brillouin zone ($k_xk_y$-plane),
and this cancels with the factor $M_1M_2$ in the integrand, giving an invariant independent of the unit cell choice.
Physically, $\nu_{ij}$ corresponds to the number of the electrons passing through 
the side of the parallelogram spanned by $\Vec{a}^{ij}_1$ and $\Vec{a}^{ij}_2$ as argued in Sec.\ \ref{sec_pump_2D},
which does not depend on the commensurability of the lattice periods.

\begin{figure*}
\centering 
\includegraphics[width=1.\linewidth]{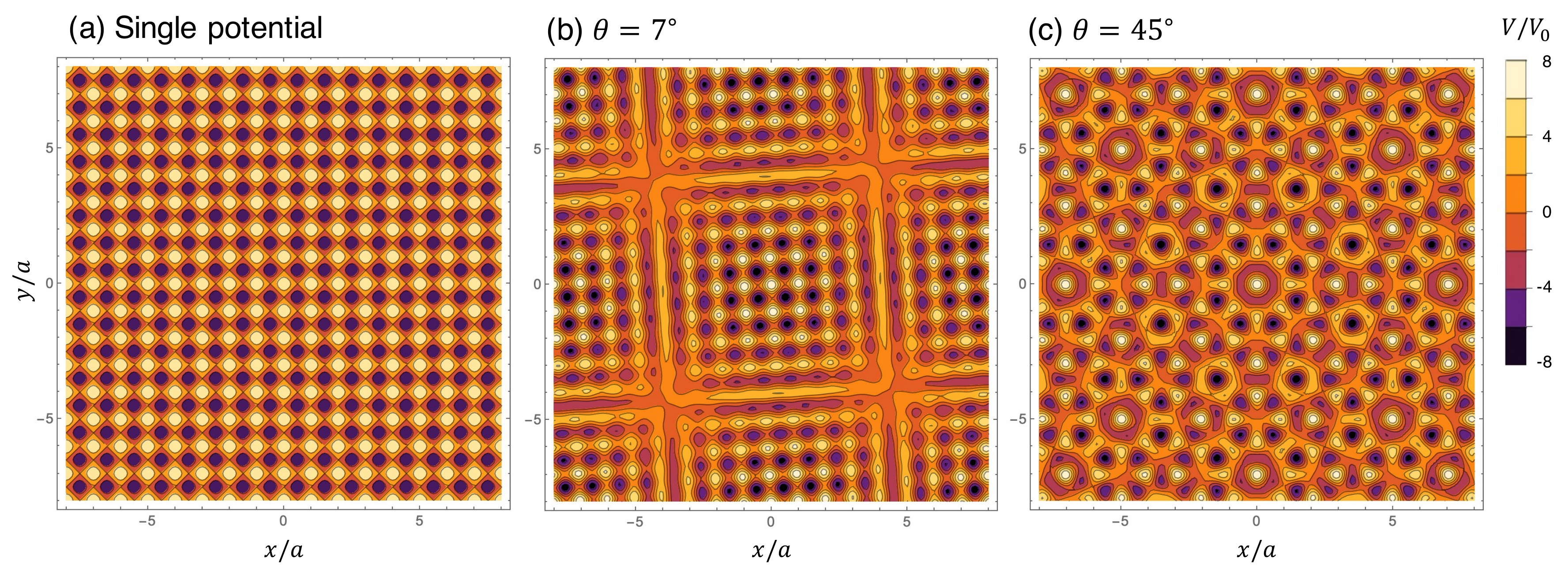}
\caption{Contour plots of (a) single square potential, and (b) twisted double square potential with $\theta = 7^\circ$ and (c) $\theta = 45^\circ$
[Eq.\ (\ref{eq_sq_potential})].}
\label{fig_twisted_square_potential}
\end{figure*}

\begin{figure*}
\centering
\includegraphics[width=1.\linewidth]{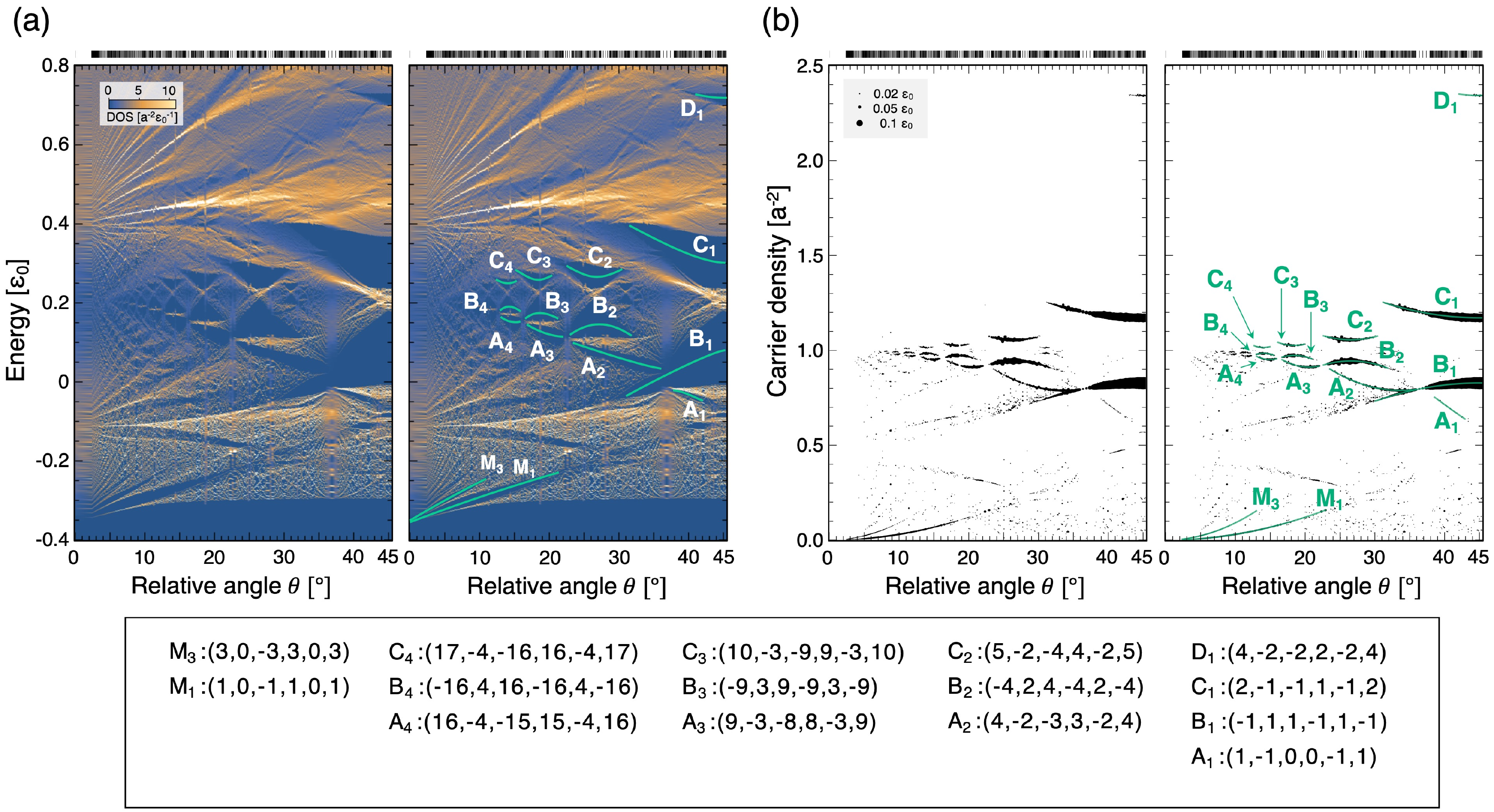}
\caption{Similar plots to Fig.\ \ref{fig_spectrum_triangular}
calculated for the twisted double square potential.
}
\label{fig_spectrum_square}
\end{figure*}

\section{Twisted square potentials}
\label{sec_twisted_sq}

We calculate the energy spectrum and the zone quantum numbers for 
a twisted double square potential. The Hamiltonian is given by Eq.\ (\ref{eq_H_2D}) with
\begin{equation}
V^\lambda(\Vec{r}) = 2V_0 \sum_{\mu=1}^2 \cos[\Vec{b}^\lambda_\mu \cdot (\Vec{r}-\Vec{r}_0^\lambda)],
\label{eq_sq_potential}
\end{equation}
where
\begin{align}
&\Vec{b}^\alpha_1 = \frac{2\pi}{a}
\begin{pmatrix}
1 \\
0
\end{pmatrix},
\quad
\Vec{b}^\alpha_2 = \frac{2\pi}{a}
\begin{pmatrix}
0 \\
1
\end{pmatrix},
\nonumber\\
&\Vec{b}^\beta_\mu= R(\theta) \,\, \Vec{b}^\alpha_\mu.
\end{align}

The corresponding primitive lattice vectors are
\begin{align}
&\Vec{a}^\alpha_1 = a
\begin{pmatrix}
1 \\
0
\end{pmatrix},
\quad
\Vec{a}^\alpha_2 = a
\begin{pmatrix}
0\\
1
\end{pmatrix},
\nonumber\\
&\Vec{a}^\beta_\mu= R(\theta) \,\, \Vec{a}^\alpha_\mu.
\end{align}
The potential profile is presented in Fig.\ \ref{fig_twisted_square_potential},
for (a) single potential,  (b) double potential with $\theta = 7^\circ$ and (c) $\theta = 45^\circ$.
The system (c) is a quasicrystal with 8-fold rotational symmetry.
The potential amplitude is taken as $V_0 = 0.213\varepsilon_0$,
where $\varepsilon_0 = \hbar^2/(2ma^2)$.

\begin{figure*}
\centering
\includegraphics[width=0.9\linewidth]{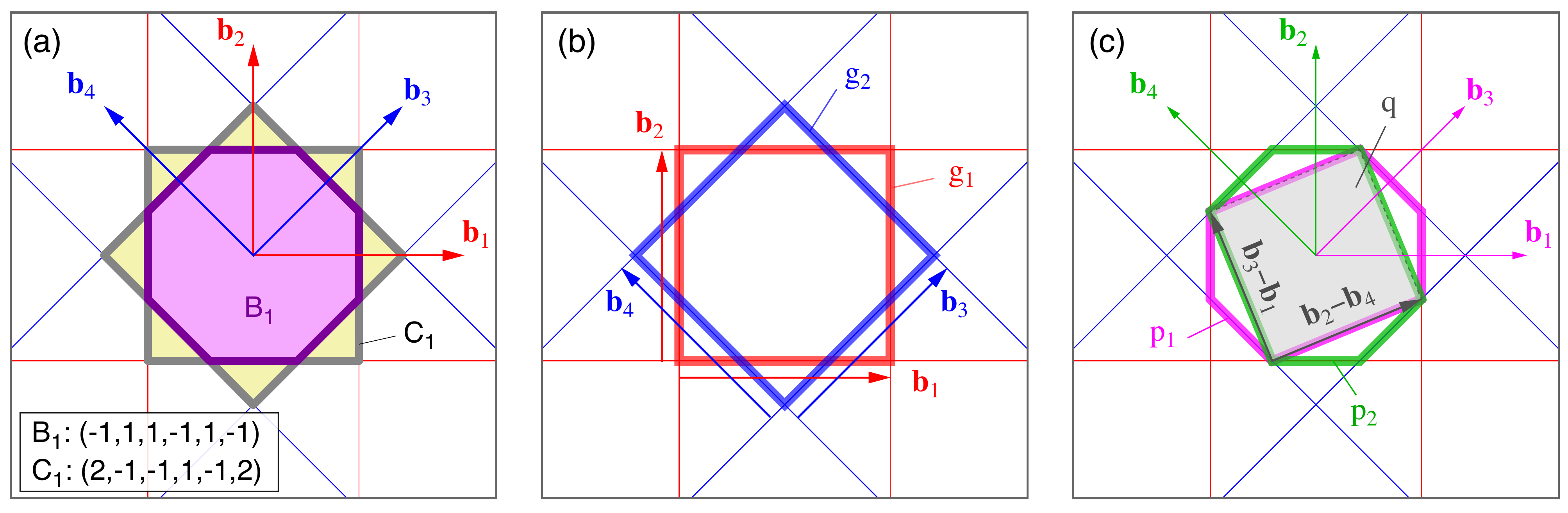}
\caption{(a) The qBZs of $B_1$ and  $C_1$ in the twisted double square potential with $\theta=45^\circ$.
(b)(c) Decomposition of the qBZ into the primitive Brillouin zones.
}
\label{fig_qBZ_square}
\end{figure*}

Figure \ref{fig_spectrum_square} is a set of plots similar to Fig.\ \ref{fig_spectrum_triangular},
calculated for the twisted square potential.
The zone quantum numbers are presented in the bottom of the figure.
We see some recursive gaps labelled by $A_n,B_n,C_n$.
We show the qBZs of the gap $B_1$ and $C_1$ at $\theta=45^\circ$ in Fig. \ref{fig_qBZ_square}(a).
The decomposition into the primitive Brillouin zone is illustrated in the right two panels.
The area of $B_1$ is given by $S^*(B_1) = (p_1+p_2) - q = S^*_{13} + S^*_{24} - (S^*_{23} + S^*_{12} + S^*_{34} - S^*_{14})$
which correctly gives the zone quantum numbers $(-1,1,1,-1,1,-1)$.
The area of $C_1$ is $S^*(C_1) = (g_1+g_2) - S^*(B_1)$, giving $(2,-1,-1,1,-1,2)$.

\section{Charge pumping formula from the infinitesimal potential limit}
\label{sec_berry}

Here we present an alternative method to derive the relation of the charge pumping to the electron density, Eq.\ (\ref{eq_P_vs_ne_2D}),
by integrating the Berry curvature in the infinitesimal potential limit. 
First, let us consider a 1D system with a single periodic potential,
\begin{equation}
H(\phi) = \frac{p^2}{2m} + V\Bigl(x -\frac{\phi}{2\pi}a\Bigr),
\label{eq_H_1D_single}
\end{equation}
where $V(x)$ is a periodic potential with the period of $a=2\pi/b$, and the phase $\phi$ describes sliding of the potential.
If we write the periodic potential in a Fourier series as
$V(x) = \sum_m V_m e^{imbx}$,
the translated potential is expressed as 
\begin{equation}
V\Bigl(x -\frac{\phi}{2\pi}a\Bigr) = \sum_m V_m e^{- i m\phi}  e^{imbx}.
\label{eq_V_slide_1D_single}
\end{equation}
The electric polarization can be calculate by 
\begin{equation}
P(\phi) =\sum_{n \in \rm occ.}
 \int_{-b/2}^{b/2}\frac{dk}{2\pi}
\,\,
i\langle u_{nk}(\phi) | \frac{\partial}{\partial k} | u_{nk}(\phi)\rangle
\label{eq_P}
\end{equation} 
where $u_{nk}(\phi)$ is the Bloch eigen state of the $n$-th band in the Hamiltonian at phase shift $\phi$,
and occ. represents the occupied bands below the Fermi energy.
The charge transport during a single sliding process is given by $\Delta P= \int_0^{2\pi} d\phi (\partial P / \partial \phi)$.
It is expressed as the Chern number on $(k,\phi)$ space,
\begin{equation}
\Delta P = \sum_{n \in \mathrm{occ.}} 
\int_0^{2\pi} d\phi
 \int_{-b/2}^{b/2} \frac{d k}{2\pi} 
F_n(k,\phi)
\label{eq_C}
\end{equation}
where $F_n(k,\phi)$ is the Berry curvature defined by
\begin{align}
&F_n(k,\phi) =\partial_1 a_n^{(2)} - \partial_2 a_n^{(1)}, \nonumber\\
&a_n^{(i)}(k,\phi) =-i  \langle u_{nk}(\phi)| \partial_i | u_{nk}(\phi) \rangle,
\end{align}
and $\partial_1 = \partial/\partial k$ and $\partial_2 = \partial/\partial \phi$.

\begin{figure}
\centering
\includegraphics[width=0.8\linewidth]{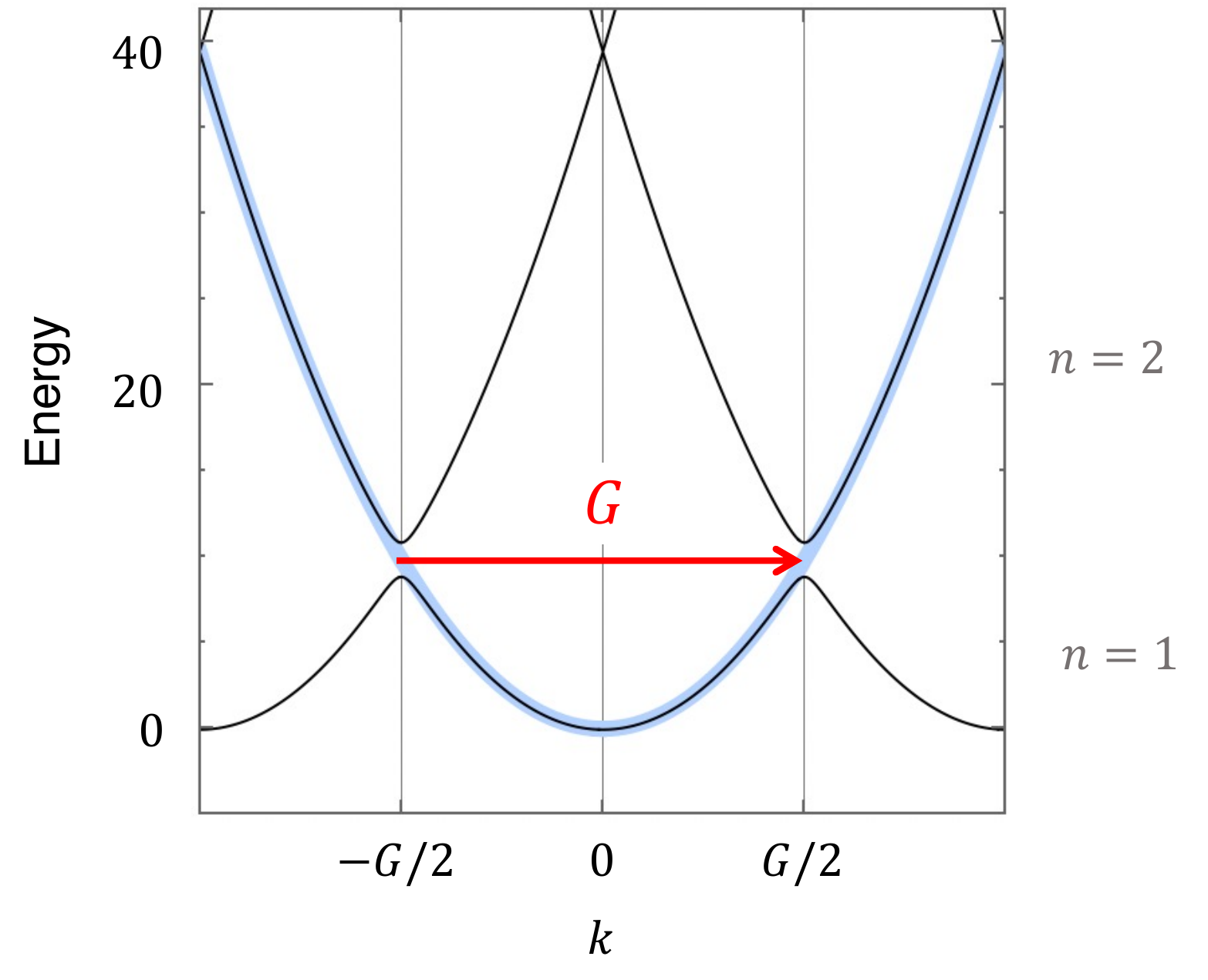}
\caption{Schematics of the energy band of a 1D periodic system in an infinitesimal potential limit.
When the periodic potential $V$ is slightly increased from zero, 
energy gaps open at $k=\pm mb/2$ ($m=1,2,3,\cdots$) in the original parabolic band of free electron (blue thin curve).
}
\label{fig_1D_band}
\end{figure}

The $\Delta P$ can be easily calculated by considering
 an infinitesimal potential limit $V(x)\to 0$.
When $V$ is slightly increased from zero, 
energy gaps open at $k=\pm mb/2$ ($m=1,2,3,\cdots$) in the original parabolic band of free electron as illustrated in Fig.\ \ref{fig_1D_band}.
Let us consider the eigenstates of the first band, $u_{1,k}$ in the first Brillouin zone $-b/2 \leq k \leq b/2$.
It is written as $|u_{1,k}\rangle = \sum_m c_m e^{i(k + mb)x}$, and we fix the global phase such that $c_0$ is real.
At the edges of the Brillouin zone, we have
\begin{align}
&|u_{1,b/2}(\phi)\rangle  \approx  \frac{1}{\sqrt{2}} \, e^{i(b/2)x} +  \frac{e^{i \phi}}{\sqrt{2}} \, e^{i(-b/2)x}
\nonumber\\
&|u_{1,-b/2}(\phi)\rangle \approx  \frac{e^{-i \phi}}{\sqrt{2}} \, e^{i(b/2)x} +  \frac{1}{\sqrt{2}} \, e^{i(-b/2)x}.
\end{align}
They are the same states but differ in the global phase factor by $e^{i \phi}$.
By using the Stokes theorem to Eq.\ (\ref{eq_C}), the Chern number of the first band is just given by
\begin{equation}
\Delta P = \frac{1}{2\pi}\int_{0}^{2\pi} d \phi
[a^{(2)}_1(b/2, \phi) - a^{(2)}_1(-b/2, \phi) ]
\end{equation}
Since $|u_{1,b/2}(\phi)\rangle = e^{i \phi} |u_{1,-b/2}(\phi)\rangle$,
we find $a^{(2)}_1(b/2, \phi) = a^{(2)}_1(-b/2, \phi) + 2\pi$, and $\Delta P=1$ is concluded.
In increasing the potential $V$, the Chern numbers do not change as long as the gap remains opening. 

The charge pumping for $m$-th gap at $k=\pm mb/2$ can also be calculated in the same manner.
We note that any perturbational processes to open the $m$-th gap
share the same $\phi$-dependent phase factor $e^{- i m \phi}$.
For instance, the first order process of $m$-th harmonics has an amplitude of $V_m e^{- i m \phi}$
and  the $m$-th order process of the first harmonics is proportional to $(V_1 e^{- i \phi})^m$.
We can integrate the Berry curvature for the 2D torus of $-mb/2 \leq k \leq mb/2$ and $0\leq \phi \leq 2\pi$
just as for the first gap.
Noting that the phase factor $e^{-i \phi}$ for the first gap is just replaced with $e^{- i m \phi}$,
we conclude $\Delta P = m$.
Here we neglected all the gaps in the occupied states below $m$-th gap,
because they do not affect the sum of the Berry curvature.

The argument also applies to a 1D doubly-periodic system of Eq.\ (\ref{eq_H_1D}).
When the potential is increased from zero, 
energy gaps open at $k=\pm (m_1 G_1 + m_2 G_2)/2$ of the original parabolic band.
The corresponding matrix element has the phase factor of $e^{- i (m_1\phi_1 + m_2 \phi_2)}$.
The charge pumping under a unit slide of the potential $V_i$ is 
calculated by integrating the Berry curvature on $(k,\phi_i)$ space, to obtain $\Delta P_i = m_i$.
It agrees with the results in Sec.\ \ref{sec_pump_1D}.

\begin{figure}
\centering
\includegraphics[width=1.\linewidth]{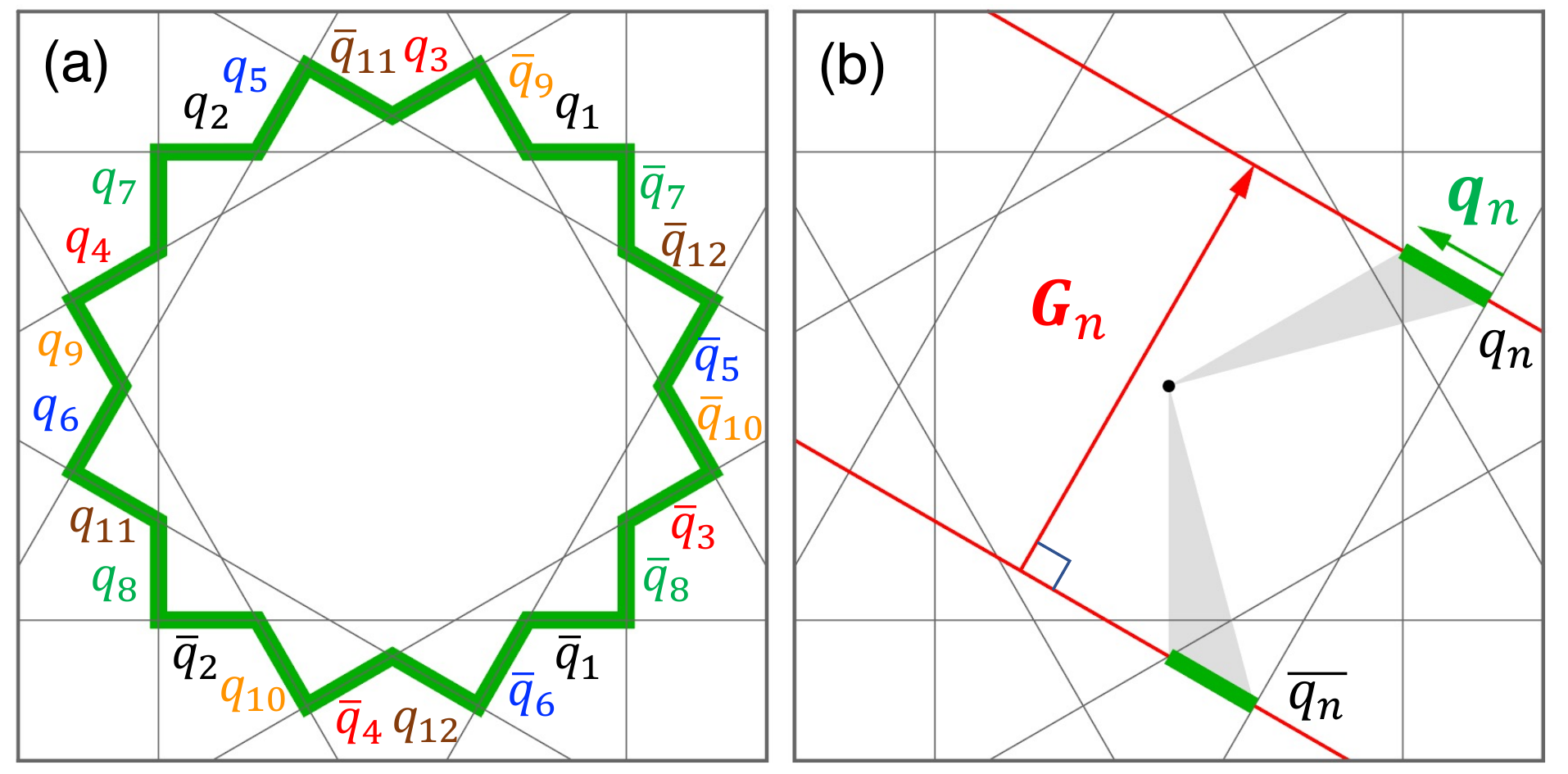}
\caption{
(a) A qBZ boundary composed of the Bragg line segments
$q_1, q_2,\cdots q_N;  \bar{q}_1, \bar{q}_2, \cdots, \bar{q}_N$,
where $q_n$ and $\bar{q}_n$ are a symmetric pair connected by the reciprocal lattice vector $\Vec{G}_n$
as shown in (b).
}
\label{fig_qBZ_indexes}
\end{figure}


The argument can be extended in a straightforward manner to a 2D doubly-periodic system, Eq.\ (\ref{eq_H_2D}).
We consider a potential sliding expressed by Eq.\ (\ref{eq_V_2D_slide}),
and calculate the change of the polarization $\Delta \Vec{P}_i$ under the process changing $\phi_i\,(i=1,2,3,4)$ from 0 to $2\pi$.
As argued in Sec.\ \ref{sec_qBZ},  each energy gap of the system 
is associated with a qBZ in the  infinitesimal potential limit.
When the potential is slightly increased from zero, the band anti-crossing occurs at the boundary of qBZ.

We can integrate the Berry phase accumulated near the gap in a similar manner to the 1D case.
Let us consider a general qBZ composed of the composite Bragg line segments
$q_1, q_2,\cdots q_N;  \bar{q}_1, \bar{q}_2, \cdots, \bar{q}_N$ as illustrated in Fig.\ \ref{fig_qBZ_indexes}(a), 
where $q_n$ and $\bar{q}_n$ are a symmetric pair connected by the reciprocal lattice vector. For illustration, we take $Q_{-1,1}$ of the $30^\circ$ case just as an example.
We take a certain pair of segments $(q_n, \bar{q}_n)$, and calculate its contribution to $\Delta \Vec{P}_i$.
Let $q_n$  be a perpendicular bisector of $\Vec{G}_n = m_{n1} \Vec{b}_1 + m_{n2}\Vec{b}_2 + m_{n3} \Vec{b}_3 + m_{n4} \Vec{b}_4$,
and define $\Vec{q}_n=(q_{nx},q_{ny})$ as a vector connecting the two ends of the segment $q_n$ as shown in Fig.\ \ref{fig_qBZ_indexes}(b).
We choose the direction of  $\Vec{q}_n$ so that $(\Vec{G}_n\times \Vec{q}_n)_z > 0$.
The matrix element associated with the gap includes the phase factor $e^{- i (m_{n1}\phi_1 + m_{n2} \phi_2 + m_{n3}\phi_3 + m_{n4} \phi_4)}$.
Let us consider a process to change $\phi_i$ from 0 to $2\pi$.
Following the discussion for the 1D case, the corresponding polarization change along $x$ at fixed $k_y$ is equal to $m_{ni}$.
The total polarization change contributed from the segments $(q_n,\bar{q}_n)$ is obtained by integrating it along $k_y$, giving $\Delta P^{(x)}_i = m_{ni}  q_{ny}/(2\pi)$.
For the $y$ component, similarly, we have $\Delta P^{(y)}_i = -m_{ni} q_{nx}/(2\pi)$.
Since  $\Vec{q}_n$ is perpendicular to $\Vec{G}_n$, it is written in a vector form as 
$\Delta \Vec{P}_i = m_{ni}  |\Vec{q}_n| (\Vec{G}_n/|\Vec{G}_n|)/(2\pi)$. By taking a summation over all the segments of the qBZ, we obtain
\begin{equation}
\Delta \Vec{P}_i = \sum_n \frac{1}{2\pi}m_{ni}  |\Vec{q}_n| \frac{\Vec{G}_n}{|\Vec{G}_n|}.
\label{eq_Delta_P_from_e_i}
\end{equation}

We can relate the polarization change Eq.\ (\ref{eq_Delta_P_from_e_i}) to the derivative of the electronic density as follows.
The electronic density $n_e$ for below the gap is given by the area of the qBZ divided by $(2\pi)^2$.
When $\Vec{b}_i$ is changed by $\delta\Vec{b}_i$,
$\Vec{G}_n$ changes by $\delta\Vec{G}_n = m_{ni} \delta\Vec{b}_i$,
and it contributes to the change of the qBZ area 
by $|\Vec{q}_n| \delta|\Vec{G}_n|= |\Vec{q}_n| \Vec{G}_n \cdot (m_{ni} \delta\Vec{b}_i)/|\Vec{G}_n|$.
As a result, the change of the electron density becomes
\begin{align}
\delta n_e
&=  
\sum_n \frac{1}{(2\pi)^2} |\Vec{q}_n|
\frac{\Vec{G} \cdot (m_{ni} \delta\Vec{b}_i)}{|\Vec{G}_n|}
\nonumber\\
&= \frac{1}{2\pi} \Delta\Vec{P}_i  \cdot \delta\Vec{b}_i,
\end{align}
where we used Eq.\ (\ref{eq_Delta_P_from_e_i}).
This immediately gives
\begin{equation}
\frac{\partial n_e}{\partial \Vec{b}_i} 
=  \frac{1}{2\pi} \Delta\Vec{P}_i,
\end{equation}
which is Eq.\ (\ref{eq_P_vs_ne_2D}).


\bibliography{Q-crystal_pump}

\end{document}